\documentclass{article}
\usepackage[latin1]{inputenc}
\usepackage[T1]{fontenc}
\usepackage{bm}
\usepackage{amsmath,amssymb,amsfonts,amscd,amsthm}
\usepackage{mathtools}
\usepackage{tabularx}
\usepackage{graphicx,color}
\usepackage{dsfont}
\usepackage{lipsum,multicol}
\usepackage{pifont}
\usepackage{nopageno}
\usepackage{appendix}
\usepackage{tabulary}
\usepackage{booktabs}
\usepackage{collcell}
\usepackage{array}
\usepackage{algorithm,algpseudocode}
\usepackage{hyperref}
\usepackage{multirow}
\usepackage[margin=1in]{geometry}
\usepackage[font=scriptsize,labelfont=bf,margin=0.1in]{caption}

\usepackage{authblk}

\usepackage{xr}
\title{Prediction and Quantification of Individual Athletic Performance}

\author[1,2]{Duncan A.J.~Blythe
\thanks{\url{duncan.blythe@bccn-berlin.de}}
}

\author[3]{
Franz J.~Kir\'{a}ly
\thanks{\url{f.kiraly@ucl.ac.uk}}
}

\affil[1]{
Bernstein Center for Computational Neuroscience Berlin,\newline
Humboldt-Universit\"at zu Berlin,
Philippstr. 13, Haus 6,
10115 Berlin, Germany
}

\affil[2]{Department of Electrical Engineering and Computer Science,\newline
Technische Universit\"at Berlin,
Stra\ss e des 17. Juni 135,
10623 Berlin, Germany
}

\affil[3]{
Department of Statistical Science,
University College London,\newline
Gower Street,
London WC1E 6BT, United Kingdom
}
\date{}

\begin{document}

\maketitle

\let\thefootnote\relax\footnote{*,$^\dag$ both authors contributed equally}

\begin{abstract}
We provide scientific foundations for athletic performance prediction on an individual level, exposing the phenomenology of individual athletic running performance in the form of a low-rank model dominated by an individual power law. We present, evaluate, and compare a selection of methods for prediction of individual running performance, including our own, \emph{local matrix completion} (LMC), which we show to perform best. We also show that many documented phenomena in quantitative sports science, such as the form of scoring tables, the success of existing prediction methods including Riegel's formula, the Purdy points scheme, the power law for world records performances and the broken power law for world record speeds may be explained on the basis of our findings in a unified way.
\end{abstract}

{\bf Note.} \emph{This manuscript is work in progress and has not yet been reviewed by an independent panel of experts. Once the manuscript has been reviewed and accepted by such a panel, this note will be removed. Until then, we would advise the reader to treat the presented results as preliminary and not to understand or present our findings as scientific fact but merely as a basis for scientific discussion.}

\section*{An overview on athletic performance prediction and our contributions}
Performance prediction and modeling are cornerstones of sports medicine, essential in training and assessment of athletes with implications beyond sport, for example in the understanding of aging, muscle physiology, and the study of the cardiovascular system.
Existing research on athletic performance focuses either on (A) explaining world records~\cite{lietzke1954analytical, henry1955prediction, riegel1980athletic, katz1994fractal, savaglio2000human, garcia2005middle}, (B) equivalent scoring~\cite{purdy1974computer,purdy1976computer}, or (C) modelling of individual physiology~\cite{hill1924muscular,billat1996significance,wasserman1973anaerobic,noakes1990peak,billat1996use,keller1973ia, peronnet1989mathematical,van1994optimisation}.
Currently, however, there is no parsimonious model which is able to simultaneously explain individual physiology (C) and collective performance (A,B).

We present such a model, a non-linear low-rank model derived from a database of UK athletes. It levers an individual power law which explains the power laws known to apply to world records, and which allows us to derive athlete-individual training parameters from prior performances data. Performance predictions obtained using our approach are the most accurate to date, with an average prediction error of under 4 minutes (2\% rel.MAE and 3\% rel.RMSE out-of-sample, see Tables~\ref{tab:rrmse_time},~\ref{tab:rmae_time} and appendix S.I.b) for elite performances.
We anticipate that our framework will allow us to leverage existing insights in the study of world record performances and sports medicine for an improved understanding of human physiology.

Our work builds on the three major research strands in prediction and modeling of running performance, which we briefly summarize:

{\bf (A) Power law models of performance} posit a power law dependence $t = c\cdot s^\alpha$ between the duration of the distance run $t$ and the distance $s$, for constants $c$ and $\alpha$.
Power law models have been known to describe world record performances across sports for over a century~\cite{kennelly1906approximate}, and have been applied extensively to running performance~\cite{lietzke1954analytical, henry1955prediction, riegel1980athletic, katz1994fractal, savaglio2000human, garcia2005middle}. These power laws have been applied by practitioners for prediction: the Riegel formula~\cite{riegel1977time} predicts performance by fitting $c$ to each athlete and fixing $\alpha =1.06$ (derived from world-record performances). The power law approach has the benefit of \emph{modelling} performances in a scientifically parsimonious way.

{\bf (B) Scoring tables}, such as those of the international association of athletics federations (IAAF), render performances over disparate distances comparable by presenting them on a single scale. These tables have been published by sports associations for almost a century~\cite{purdy1974history}
and catalogue, rather than model, performances of equivalent standard. Performance predictions may be obtained from scoring tables by forecasting a time with the same score as an existing attempt, as implemented in the popular Purdy Points scheme~\cite{purdy1974computer,purdy1976computer}. The scoring table approach has the benefit of \emph{describing} performances in an empirically accurate way.

{\bf (C) Explicit modeling of performance related physiology} is an active subfield of sports science. Several physiological parameters are known to be related to athletic performance; these include maximal oxygen uptake (\.VO$_2$-max) and critical speed (speed at \.{V}O$_2$-max) \cite{hill1924muscular,billat1996significance}, blood lactate concentration, and the anaerobic threshold~\cite{wasserman1973anaerobic,bosquet2002methods}.
Physiological parameters may be used (C.i) to make direct predictions when clinical measurements are available~\cite{noakes1990peak,billat1996use,bundle2003high}, or (C.ii) to obtain theoretical models describing physiological processes~\cite{keller1973ia, peronnet1989mathematical,van1994optimisation,di2003factors}. These approaches have the benefit of \emph{explaining} performances physiologically.

All three approaches (A),(B),(C) have appealing properties, as explained above, but none provides a complete treatment of athletic performance prediction:
(A) individual performances do not follow the parsimonious power law perfectly; (B) the empirically accurate scoring tables do not provide a simple interpretable relationship.
\emph{Neither} (A) nor (B) can deal with the fact that athletes may differ from one another in multiple ways.
The clinical measurements in (C.i) are informative but usually available only for a few select athletes, typically at most a few dozen (as opposed to the 164,746 considered in our study). The interpretable models in (C.ii) are usually designed not with the aim of predicting performance but to explain physiology or to estimate physiological parameters from performances; thus these methods are not directly applicable without additional work.

The approach we present unifies the desirable properties of (A),(B) and (C), while avoiding the aforementioned shortcomings.
We obtain (A) a parsimonious model for individual athletic performance that is (B) empirically derived from a large database of UK athletes. It yields the best performance predictions to date (2\% average error for elite athletes on all events, average error 3-4 min for Marathon, see Table~\ref{tab:rmae_time}) and (C) unveils hidden descriptors for individuals which we find to be related to physiological characteristics.

Our approach bases predictions on \emph{Local Matrix Completion} (LMC), a machine learning technique which posits the existence of a small number of explanatory variables which describe the performance of individual athletes.
Application of LMC to a database of athletes allows us, in a second step, to derive a parsimonious physiological model describing athletic performance of \emph{individual} athletes. We discover that a \emph{three number-summary} for each individual explains performance over the full range of distances from 100m to the Marathon. The three-number-summary relates to: (1) the endurance of an athlete, (2) the relative balance between speed and endurance, and
(3) specialization over middle distances. The first number explains most of the individual differences over distances greater than 800m, and may be interpreted as the exponent of \emph{an individual power law for each athlete}, which holds with remarkably high precision, on average. The other two numbers describe individual, non-linear corrections to this individual power law. Vitally, we show that the individual power law with its non-linear corrections reflects the data more accurately than the power law for world records.
We anticipate that individual power law and three-number summary will allow for exact quantitative assessment in the science of running and related sports.

\begin{figure}
\begin{center}
  \includegraphics[width=150mm,clip=true,trim= 5mm 20mm 5mm 20mm]{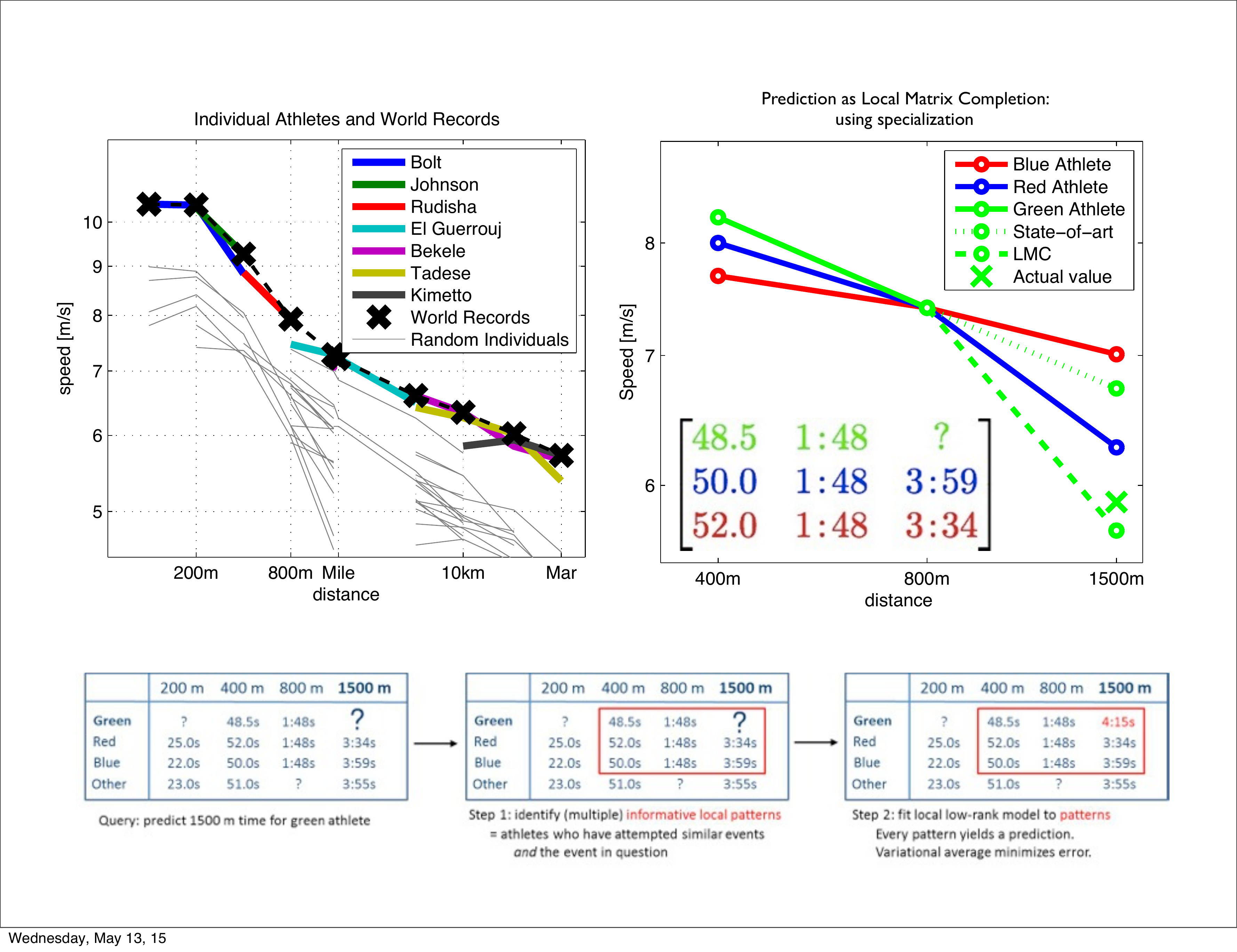} \\
\end{center}
\caption{Non-linear deviation from the power law in individuals as central phenomenon. Top left: performances of world record holders and a selection of random athletes. Curves labelled by athletes are their known best performances (y-axis) at that event (x-axis). Black crosses are world record performances. Individual performances deviate non-linearly from the world record power law. Top right: a good model should take into account specialization, illustration by example. Hypothetical performance curves of three athletes, green, red and blue are shown, the task is to predict green on 1500m from all other performances. Dotted green lines are predictions. State-of-art methods such as Riegel or Purdy predict green performance on 1500m close to blue and red; a realistic predictor for 1500m performance of green  - such as LMC - will predict that green is outperformed by red and blue on 1500m; since blue and red being worse on 400m indicates that out of the three athletes, green specializes most on shorter distances. Bottom: using local matrix completion as a mathematical prediction principle by filling in an entry in a $(3\times 3)$ sub-pattern. Schematic illustration of the algorithm.
\label{fig:illustration}}
\end{figure}

\section*{Local Matrix Completion and the Low-Rank Model}

It is well known that world records over distinct distances are held by distinct athletes---no one single athlete holds all running world records. Since world record data obey an approximate power law, this implies that the individual performance of each athlete deviates from this power law.
The left top panel of Figure~\ref{fig:illustration} displays world records and the corresponding individual performances of world record holders in logarithmic coordinates---an exact power law would follow a straight line. The world records align closely to a straight line, while individuals deviate non-linearly. Notable is also the kink in the world records which makes them deviate from an exact straight line, yielding a ``broken power law'' for world records~\cite{savaglio2000human}.

Any model for individual performances must model this individual, non-linear variation - and will, optimally, explain the broken power law observed for world records as an epiphenomenon of such variation over individuals. In following paragraphs we explain how the LMC scheme captures individual variation in a typical scenario.

Consider three athletes (taken from the data base) as shown in the right top panel of Figure~\ref{fig:illustration}. The 1500m performance of the green athlete is not known and is to be predicted. All three athletes, green, blue and red, have similar performance on 800m. Any classical method for performance prediction which only takes that information into account will predict that green performs similarly on 1500m to the blue and the red, e.g.~somewhere in-between. However, this is unrealistic, since it does not take into account event specialization: looking at the 400m performance, one can see that the red athlete is slowest over short distances, followed by the blue and then by the green whose relative speed surpasses the remaining athletes over longer distances. Using this additional information leads to the more realistic prediction that the the green athlete will be out-performed by red and blue on 1500m. Supplementary analysis (S.IV) validates that the phenomenon presented in the example is prevalent throughout the data set.

LMC is a quantitative method for taking into account this event specialization. A schematic overview of the simplest variant is displayed in the bottom panel of Figure~\ref{fig:illustration}: to predict an event for an athlete (figure: 1500m for green) we find a 3-by-3-pattern of performances, denoted by $A$, with exactly one missing entry - this means the two other athletes (figure: red and blue) have attempted similar events and have data available. Explanation of the green athlete's curve by the red and the blue is mathematically modelled by demanding that the data of the green athlete is given as a weighted sum of the data of the red and the blue; i.e., more mathematically, the green row is a linear combination of the blue and the red row. By a classical result in matrix algebra, the green row is a linear combination of red and blue whenever the \emph{determinant} of $A$, a polynomial function in the entries of $A$, vanishes; i.e., $\det(A) = 0$.

A prediction is made by solving the equation $\det(A) = 0$ for ``?''. To increase accuracy, candidate solutions from multiple 3-by-3-patterns (obtained from many triples of athletes) are averaged in a way that minimizes the expected error in approximation. We will consider variants of the algorithm which use $n$-by-$n$-patterns, $n$ corresponding to the complexity of the model (we later show $n=4$ to be optimal). See the methods appendix for an exact description of the algorithm used.

The LMC prediction scheme is an instance of the more general local low-rank matrix completion framework introduced in~\cite{kiraly15matrixcompletion}, here applied to performances in the form of a numerical table (or matrix) with columns corresponding to events and rows to athletes. The cited framework is the first matrix completion algorithm which allows prediction of single missing entries as opposed to all entries. While matrix completion has proved vital in predicting consumer behaviour and recommender systems, we find that existing approaches which predict all entries at once cannot cope with the non-standard distribution of missingness and the noise associated with performance prediction in the same way as LMC can (see findings and supplement S.II.a). See the methods appendix for more details of the method and an exact description.

In a second step, we use the LMC scheme to fill in all missing performances (over all events considered---100m, 200m etc.) and obtain a parsimonious low-rank model that explains individual running times $t$ in terms of distance $s$ by:
\begin{align}
\log t = \lambda_1 f_1(s) + \lambda_2 f_2(s) + \dots + \lambda_r f_r(s) \label{eq:model},
\end{align}
with components $f_1$, $f_2$, $\dots$ that are universal over athletes, and coefficients $\lambda_1$, $\lambda_2$, $\dots$, $\lambda_r$ which summarize the athlete under consideration.
The number of components and coefficients $r$ is known as the \emph{rank} of the model and measures its complexity; when considering the data in matrix form, $r$ translates to
\emph{matrix rank}.
The Riegel power law is a very special case, demanding that $\lambda_1 = 1.06$ for every athlete, $ f_1(s) = \text{log}~s$ and  $\lambda_2 f_2(s) = c$ for a constant $c$ depending on the athlete. Our analyses will show that the best model has rank $r = 3$ (meaning above we consider patterns or matrices of size $n \times n = 4\time 4$ since above $n=r+1$). This means that the model has $r=$ \emph{three} universal components $f_1(s), f_2(s), f_3(s)$, and every athlete is described by their individual three-coefficient-summary $\lambda_1,\lambda_2,\lambda_3$. Remarkably, we find that $f_1(s) = \text{log}~s$, yielding an individual power law; the corresponding coefficient $\lambda_1$ thus has the natural interpretation as an individual power law exponent.

We remark that first filling in the entries with LMC and only then fitting the model is crucial due to data which is non-uniformly missing (see supplement S.II.a). More details on our methodology can be found in the methods appendix.

\subsection*{Data Set, Analyses and Model Validation}

The basis for our analyses is the online database {\tt www.thepowerof10.info}, which catalogues British individuals' performances achieved in officially ratified athletics competitions since 1954.
The excerpt we consider dates from August 3, 2013. It contains (after error removal) records of 164,746 individuals of both genders, ranging from the amateur to the elite, young to old, comprising a total of 1,417,432 individual performances over 10 different distances: 100m, 200m, 400m, 800m, 1500m, the Mile, 5km, 10km, Half-Marathon, Marathon (42,195m). All British records over the distances considered are contained in the dataset; the 95th percentile for the 100m, 1500m and Marathon are 15.9, 6:06.5 and 6:15:34, respectively.
As performances for the two genders distribute differently, we present only results on the 101,775 male athletes in the main corpus of the manuscript; female athletes and subgroup analyses are considered in the supplementary results.
The data set is available upon request, subject to approval by British Athletics. Full code of our analyses can be obtained from [download link will be provided here after acceptance of the manuscript].

Adhering to state-of-the-art statistical practice (see ~\cite{efron1983estimating,kohavi1995study,efron1997improvements,browne2000cross}), all prediction methods are validated \emph{out-of-sample}, i.e., by using only a subset of the data for estimation of parameters (training set) and computing the error on predictions made for a distinct subset (validation or test set). As error measures, we use the root mean squared error (RMSE) and the mean absolute error (MAE), estimated by leave-one-out validation for 1000 single performances omitted at random.

We would like to stress that out-of-sample prediction error is the correct way to evaluate the quality of \emph{prediction}, as opposed to merely reporting goodness-of-fit in-sample; since outputting an estimate for an instance that the method has already seen does not qualify as prediction.

More details on the data set and our validation setup can be found in the supplementary material.

\section*{Findings on the UK athletes data set}

{\bf (I) Prediction accuracy.} We evaluate prediction accuracy of ten methods, including our proposed method, LMC. We include, as
{\it naive baselines:} (1.a) imputing the event mean, (1.b) imputing the average of the $k$-nearest neighbours; as representative of the {\it state-of-the-art in quantitative sports science:} (2.a) the Riegel formula, (2.b) a power-law predictor with exponent estimated from the data, which is the same for all athletes, (2.c) a power-law predictor with exponent estimated from the data, one exponent per athlete, (2.d) the Purdy points scheme \cite{purdy1974computer}; as representatives for the {\it state-of-the-art in matrix completion:} (3.a) imputation by expectation maximization on a multivariate Gaussian \cite{dempster1977maximum} (3.b) nuclear norm minimization~\cite{candes2009exact,candes2010power}.\\
We instantiate our {\it low-rank local matrix completion} (LMC) in two variants: (4.a) rank 1, and (4.b) rank 2.

Methods (1.a), (1.b), (2.a), (2.b), (2.d), (4.a) require at least one observed performance per athlete, methods (2.c), (4.b) require at least two observed performances in distinct events. Methods (3.a), (3.b) will return a result for any number of observed performances (including zero). Prediction accuracy is therefore measured by evaluating the RMSE and MAE out-of-sample on the athletes who have attempted at least three distances, so that the two necessary performances remain when one is removed for leave-one-out validation. Prediction is further restricted to the best 95-percentile of athletes (measured by performance in the best event) to reduce the effect of outliers. Whenever the method demands that the predicting events need to be specified, the events which are closest in log-distance to the event to be predicted are taken. The accuracy of predicting time (normalized w.r.t.~the event mean), log-time, and speed are measured.
We repeat this validation setup for the year of best performance and a random calendar year. Moreover, for completeness and comparison we
treat 2 additional cases: the top 25\% of athletes and athletes who have attempted at least 4 events, each in log time.
More details on methods and validation are presented in the methods appendix.

The results are displayed in Table~\ref{tab:compare_methods} (RMSE) and supplementary Table~\ref{tab:compare_methods_mae} (MAE). Of all benchmarks, Purdy points (2.d) and Expectation Maximization (3.a) perform best. LMC in rank 2 substantially outperforms Purdy points and Expectation Maximization (two-sided Wilcoxon signed-rank test significant at $p \le$ 1e-4 on the validation samples of absolute prediction errors); rank 1 outperforms Purdy points on the year of best performance data ($p=$5.5e-3) for the best athletes, and is  on a par on athletes up to the 95th percentile. Both rank 1 and 2 outperform the power law models ($p\le$1e-4), the improvement in RMSE over the power-law reaches over 50\% for data from the fastest 25\% of athletes.

{\bf  (II) The rank (number of components) of the model.} Paragraph (I) establishes that LMC is the best method for prediction. LMC assumes a fixed number of prototypical athletes, viz.~the rank $r$, which is the complexity parameter of the model.
We establish the optimal rank by comparing prediction accuracy of LMC with different ranks. The rank $r$ algorithm needs $r$ attempted events for prediction, thus $r+1$ observed events are needed for validation. Table~\ref{tab:determine_rank} displays prediction accuracies for LMC ranks $r=1$ to $r=4$, on the athletes who have attempted $k > r$ events, for all $k \le 5$. The data is restricted to the top 25\% in the year of best performance in order to obtain a high signal to noise ratio. We  observe that rank 3 outperforms all other ranks, when applicable; rank 2 always outperforms rank 1 (both $p\le$1e-4).

We also find that the improvement of rank 2 over rank 1 depends on the event predicted: improvement is 26.3\% for short distances (100m,200m), 29.3\% for middle distances (400m,800m,1500m), 12.8\% for the mile to half-marathon, and 3.1\% for the Marathon (all significant at $p$=1e-3 level) (see Figure~\ref{fig:individual_events}).
These results indicate that inter-athlete variability is greater for short and middle distances than for Marathon.

\begin{figure}
\begin{center}
$\begin{array}{c c c}
\includegraphics[width=45mm,clip=true,trim= 0mm 0mm 0mm 0mm]{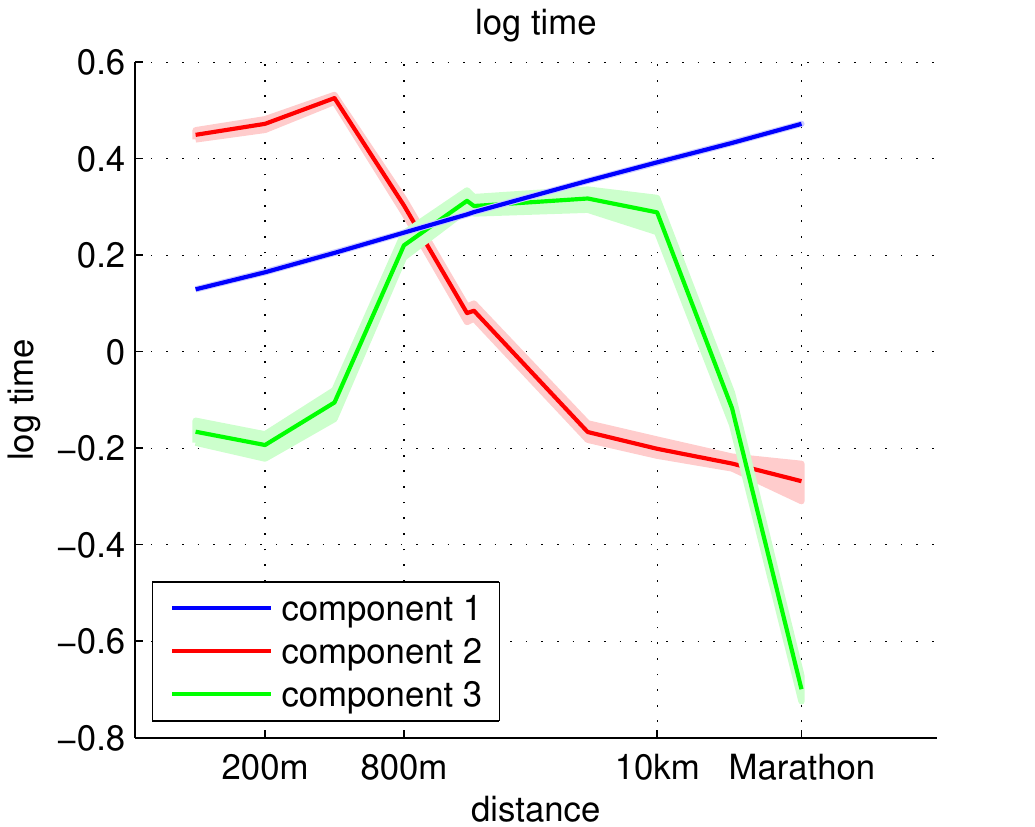} &
\includegraphics[width=45mm,clip=true,trim= 0mm 0mm 0mm 0mm]{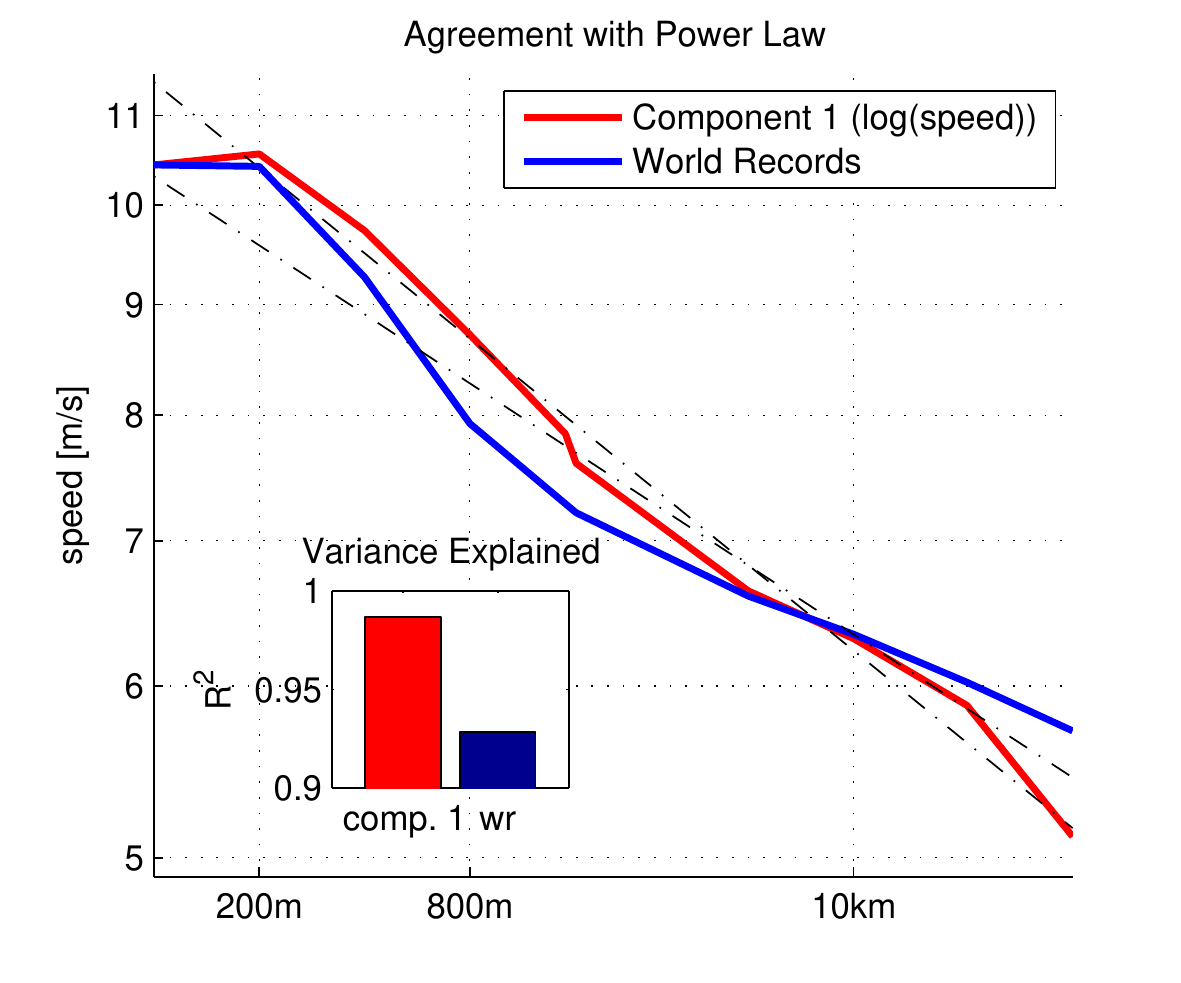} &
\includegraphics[width=45mm,clip=true,trim= 0mm 0mm 0mm 0mm]{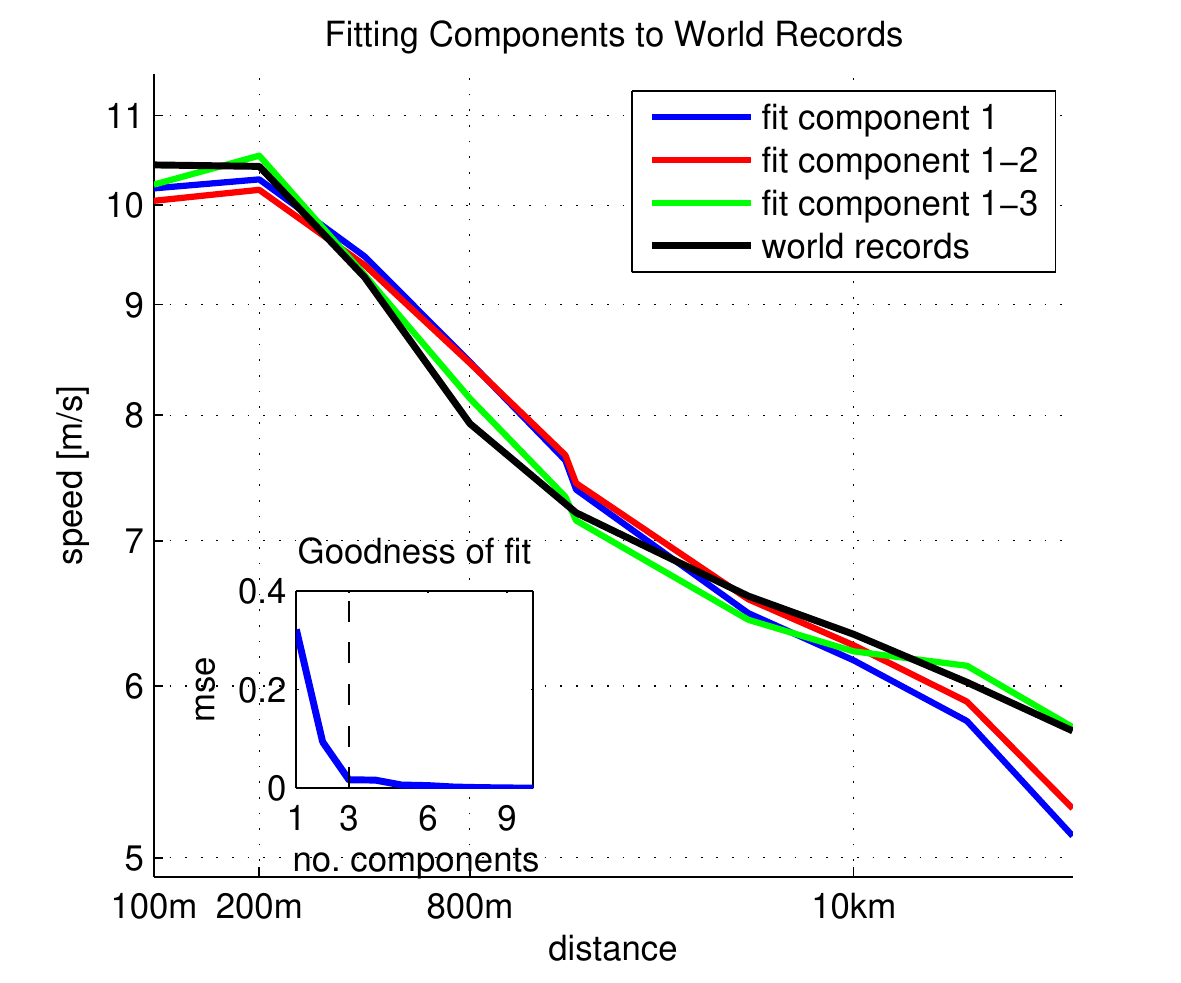}

\end{array}$
\end{center}
\caption{The three components of the low-rank model, and explanation of the world record data. Left: the components displayed (unit norm, log-time vs log-distance). Tubes around the components are one standard deviation, estimated by the bootstrap. The first component is an exact power law (straight line in log-log coordinates); the last two components are non-linear, describing transitions at around 800m and 10km. Middle: Comparison of first component and world record to the exact power law (log-speed vs log-distance). Right: Least-squares fit of rank 1-3 models to the world record data (log-speed vs log-distance).
\label{fig:svs}}
\end{figure}

{\bf (III) The three components of the model.} The findings in (II) imply that the best low-rank model assumes $3$ components. To estimate the components ($f_i$ in Equation~\eqref{eq:model}) we impute all missing entries in the data matrix of the top 25\% athletes who have attempted 4 events and compute its singular value decomposition (SVD)~\cite{golub1970singular}. From the SVD, the exact form of components can be directly obtained as the right singular vectors (in a least-squares sense, and up to scaling, see methods appendix). We obtain three components in log-time coordinates, which are displayed in the left hand panel of Figure~\ref{fig:svs}. The first component for log-time prediction is linear (i.e., $f_1(s) \propto \log s$ in Equation~\eqref{eq:model}) to a high degree of precision ($R^2 = 0.9997$) and corresponds to an \emph{individual} power law, applying distinctly to each athlete. The second and third components are non-linear; the second component decreases over short sprints and increases over the remainder, and the third component resembles a parabola with extremum positioned around the middle distances.

In speed coordinates, the first, individual power law component does not display the ``broken power law'' behaviour of the world records. Deviations from an exact line can be explained by the second and third component (Figure~\ref{fig:svs} middle).

The three components together explain the world record data and its ``broken power law'' far more accurately than a simple linear power law trend---with the rank 3 model fitting the world records almost exactly (Figure~\ref{fig:svs} right;
rank 1 component: $R^2 = 0.99$; world-record data: $R^2 = 0.93$).

\begin{figure}
\begin{center}
$\begin{array}{c}
\includegraphics[width=160mm,clip=true,trim= 5mm 70mm 5mm 60mm]{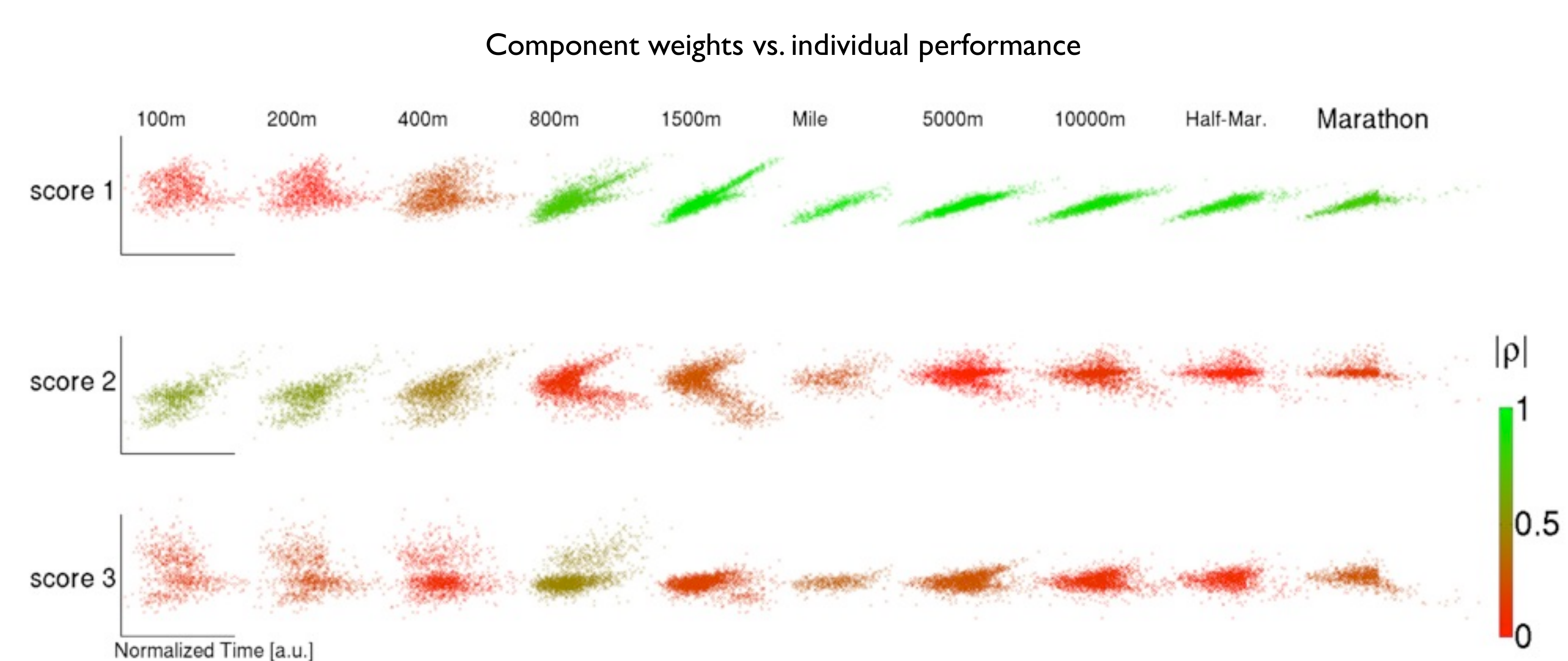}
\end{array}$
\end{center}
\caption{Matrix scatter plot of the three-number-summary vs performance. For each of the scores in the three-number-summary (rows) and each event distance (columns), the plot matrix shows: a scatter plot of performances (time) vs the coefficient score of the top 25\% (on the best event) athletes who have attempted at least 4 events. Each scatter plot in the matrix is colored on a continuous color scale according to the absolute value of the scatter sample's Spearman rank correlation (red = 0, green = 1).
\label{fig:scores}
}
\end{figure}

\begin{figure}
\begin{center}
$\begin{array}{c c}
\includegraphics[width=55mm,clip=true,trim= 0mm 0mm 0mm 0mm]{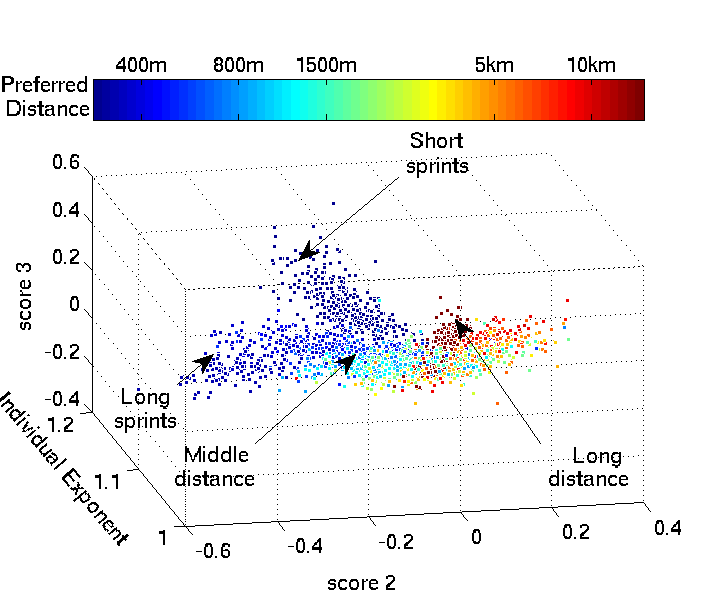} &
\includegraphics[width=55mm,clip=true,trim= 0mm 0mm 0mm 0mm]{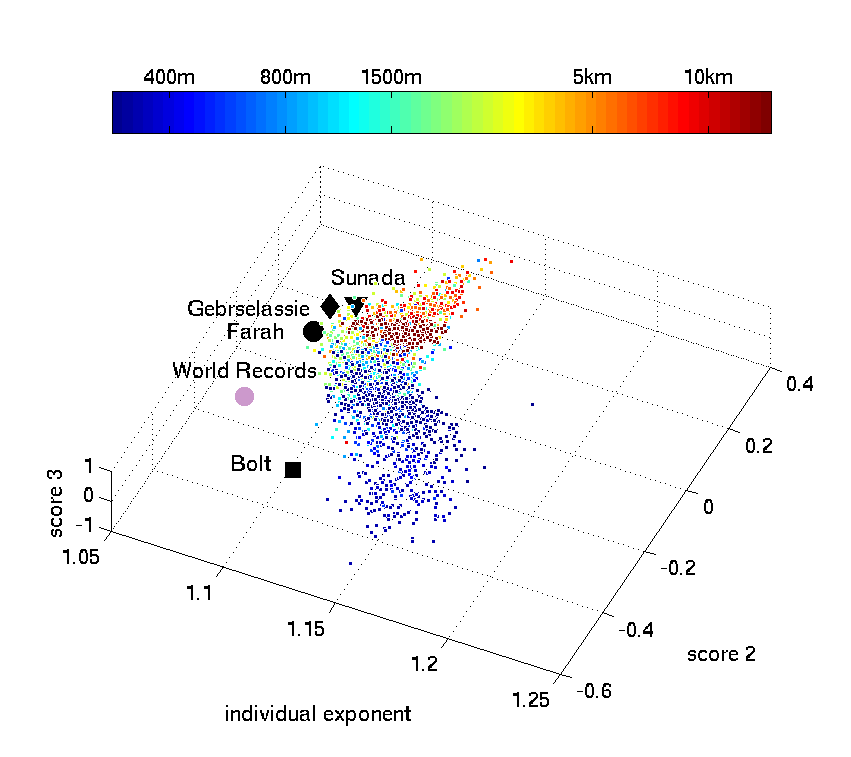}\\
\includegraphics[width=55mm,clip=true,trim= 0mm 0mm 0mm 0mm]{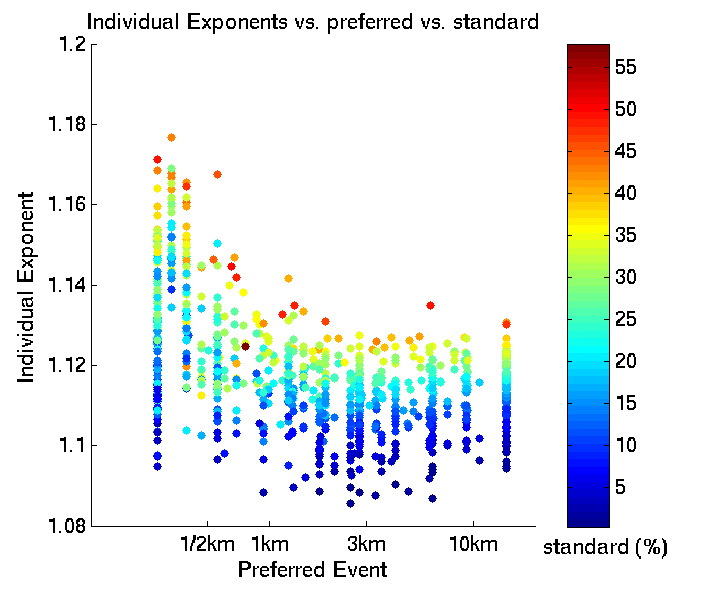} &
\includegraphics[width=55mm,clip=true,trim= 0mm 0mm 0mm 0mm]{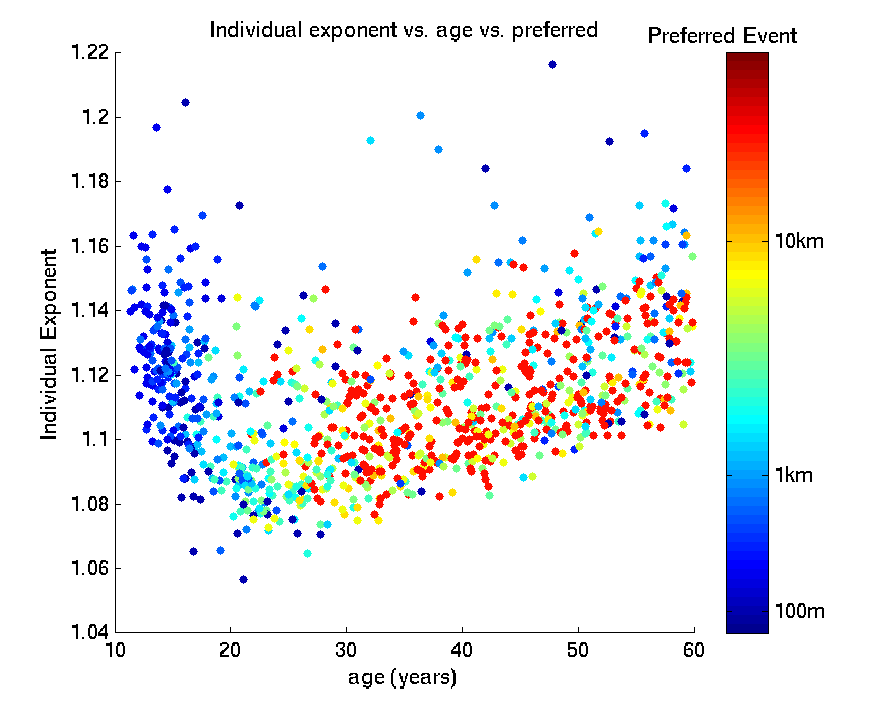}
\end{array}$
\end{center}
\caption{Scatter plots exploring the three number summary. Top left and right: 3D scatter plot of three-number-summaries of athletes in the data set, colored by preferred distance and shown from two angles. A negative value for the second score is a indicates that the athlete is a sprinter, a positive value an endurance runner. In the top right panel, the summaries of the elite athletes Usain Bolt (world record holder, 100m, 200m), Mo Farah (world beater over distances between 1500m and 10km), Haile Gabrselassie (former world record holder from 5km to Marathon) and Takahiro Sunada (100km world record holder) are shown; summaries are estimated from their personal bests. For comparison we also display the hypothetical data of an athlete
who holds all world records.
Bottom left: preferred distance vs individual exponents, color is percentile on preferred distance. Bottom right: age vs. exponent, colored by preferred distance.
\label{fig:scatter}}
\end{figure}

{\bf (IV) The three athlete-specific coefficients.} The three summary coefficients for each athlete ($\lambda_1,\lambda_2,\lambda_3$ in Equation~\eqref{eq:model}) are obtained from the entries of the left singular vectors (see methods appendix). Since all three coefficients summarize the athlete, we refer to them collectively as the \emph{three-number-summary}.
(IV.i) Figure~\ref{fig:scores} displays scatter plots and Spearman correlations between the coefficients and performance over the full range of distances. The individual exponent correlates with performance on distances greater than 800m. The second coefficient correlates positively with performance over short distances and displays a non-linear association with performance over middle distances.
The third coefficient correlates with performance over middle distances. The associations for all three coefficients are non-linear, with the notable exception of the individual exponent on distances exceeding 800m, hence the application of Spearman correlations.
(IV.ii) Figure~\ref{fig:scatter} top displays the three-number-summary for the top 95\% athletes in the data base. The athletes appear to separate into (at least) four classes, which associate with the athlete's preferred distance. A qualitative transition can be observed over middle distances. Three-number-summaries of world class athletes (not all in the UK athletics data base), computed from their personal bests, are listed in Table~\ref{tab:elite_scores}; they and also shown as highlighted points in Figure~\ref{fig:scatter} top right. The elite athletes trace a frontier around the population: all elite athletes are subject to a low individual exponent. A hypothetical athlete holding all the world records is also shown in Figure~\ref{fig:scatter} top right, obtaining an individual exponent which comes close to the world record exponent estimated by Riegel~\cite{riegel1980athletic}.
(IV.iii) Figure~\ref{fig:scatter} bottom left shows that a low individual exponent correlates positively with  performance in an athlete's preferred event. The individual exponents are higher on average (median=1.12; 5th, 95th percentiles=1.10,1.15) than the world record exponents estimated by Riegel~\cite{riegel1980athletic} (1.08 for elite athletes, 1.06 for senior athletes).
(IV.iv) Figure~\ref{fig:scatter} bottom right shows that in cross-section, the individual exponent decreases with age until 20 years, and subsequently increases.

\begin{table}
\begin{center}
\begin{tabular}{l l | c c c } 
Athlete & Specialization & Individual Exponent ($\lambda_1$) & Score 2 ($\lambda_2$)  & Score 3 ($\lambda_3$) \\ 
\hline 
Usain Bolt & Sprints & 1.11 & -0.367 & 0.0813\\ 
Mo Farah & Middle-Long & 1.08 & 0.0325 & -0.0761\\ 
Haile Gabrselassie & Long & 1.08 & 0.114 & -0.0556\\ 
Galen Rupp & Long & 1.08 & 0.104 & -0.0395\\ 
Seb Coe & Middle & 1.09 & -0.0847 & -0.0359\\ 
Takahiro Sunada & Ultra-Long & 1.09 & 0.138 & -0.00917\\ 
Paula Radcliffe & Long (Female) & 1.10 & 0.189 & 0.0254\\ 
\end{tabular}
\end{center}
\caption{Estimated three-number-summary ($\lambda_i$) in log(time) coordinates of selected elite athletes. The scores $\lambda_1$, $\lambda_2$, $\lambda_3$ are defined by Equation~\eqref{eq:model} and may be interpreted as the contribution of each component to performance for a given athlete. Since component 1 is a power-law (see the top-left of Figure~\ref{fig:svs}), $\lambda_1$ may be interpreted as the individual exponent. See the bottom right panel of Figure~\ref{fig:scatter} for a scatter plot of athletes.
\label{tab:elite_scores}
}
\end{table}

{\bf (V) Phase transitions.} We observe two transitions in behaviour between short and long distances.
The data exhibit a phase transition around 800m: the second component exhibits a kink and the third component makes a zero transition (Figure~\ref{fig:svs}); the association of the first two scores with performance shifts from the second to the first score (Figure~\ref{fig:scores}). The data also exhibits a transition around 5000m. We find that for distances shorter than 5000m, holding the event performance constant, and increasing the standard of shorter events leads
to a decrease in the predicted standard of longer events and vice versa. On the other hand for distances greater than 5000m this behaviour reverses; holding the event performance constant, and increasing the standard of shorter events
leads to an increase in the predicted standard of longer events.
See supplementary section (S.IV) for details.

{\bf (VI) Universality over subgroups.} Qualitatively and quantitatively similar results to the above can be deduced for female athletes, and subgroups stratified by age or training standard; LMC remains an accurate predictor, and the low-rank model has similar form. See supplement (S.II.b).

\section*{Discussion and Outlook}

We have presented the most accurate existing predictor for running performance---local low-rank matrix completion (finding I); its predictive power confirms the validity of a three-component model (finding II) that offers a parsimonious explanation for many known phenomena in the quantitative science of running, including answers to some of the major open questions of the field. More precisely, we establish:

{\bf The individual power law.} In log-time coordinates, the first component of our physiological model is linear with high accuracy, yielding an individual power law (finding III). This is a novel and rather surprising finding, since, although world-record performances are known to obey a power law~\cite{lietzke1954analytical, henry1955prediction, riegel1980athletic, katz1994fractal, savaglio2000human, garcia2005middle}, there is no reason to \emph{a-priori} suppose that the performance of individuals is governed by a power law. This \emph{parsimony a-posteriori} unifies (A) the parsimony of the power law with the (B) empirical correctness of scoring tables. To which extent this individual power law is exact is to be determined in future studies.

{\bf An explanation of the world record data.} The broken power law on world records can be seen as a consequence of the individual power law and the non-linearity in the second and third component (finding III) of our low-rank model. The breakage point in the world records can be explained by the differing contributions in the non-linear components of the distinct individuals holding the world records.

Thus both \emph{the power law and the broken power law on world record data can be understood as epiphenomena of the individual power law and its non-linear corrections}.

{\bf Universality of our model.} The low-rank model remains unchanged when considering different subgroups of athletes, stratified by gender, age, or calendar year; what changes is only the individual three-number-summaries (finding VI). This shows the low-rank model to be universal for running.

{\bf The three-number-summary reflects an athlete's training state.} Our predictive validation implies that the number of components of our model is three (finding II), which yields three numbers describing the training state of a given athlete (finding IV). The most important summary is \emph{the individual exponent} for the individual power law which describes overall performance (IV.iii). The second coefficient describes whether the athlete has greater endurance (positive) or speed (negative), the third describes specialization over middle distances (negative) vs short and long distances (positive). All three numbers together clearly separate the athletes into four clusters, which fall into two clusters of short-distance runners and one cluster of middle-and long-distance runners respectively (IV.i).
Our analysis provides strong evidence that the three-number-summary captures physiological and/or social/behavioural characteristics of the athletes, e.g., training state, specialization, and which distance an athlete chooses to attempt.
While the data set does not allow us to separate these potential influences or to make statements about cause and effect, we conjecture that combining the three-number-summary with specific experimental paradigms will lead to a clarification; further, we conjecture that a combination of the three-number-summary with additional data, e.g. training logs, high-frequency training measurements or clinical parameters, will lead to a better understanding of (C) existing physiological models.\\

Some novel physiological insights can be deduced from leveraging our model on the UK athletics data base:

\begin{itemize}
\itemsep0em
\item We find that the higher rank LMC predictor is most effective for the longer-sprints and middle distances, and in comparison to the rank 1 predictor; the improvement of the higher rank over the rank 1 version is lowest over the marathon distance.
This may be explained by some middle-distance runners using a high maximum velocity to coast whereas other runners use greater endurance to run closer to their maximum speed for the duration of the race; it would be interesting to check empirically whether the type of running (coasting vs endurance) is the physiological correlate to the specialization summary. If this was verified, it could imply that (presently) there is {\bf only one way to be a fast marathoner}, i.e., possessing a high level of endurance---as opposed to being able to coast relative to a high maximum speed. In any case, the low-rank model predicts that a marathoner who is not close to world class over 10km is unlikely to be a world class marathoner.

\item The {\bf phase transitions} which we observe (finding V) provide additional observational evidence for a transition in the complexity of the physiology underlying performance between long and short distances.
This finding is bolstered by the difference we observe between the increase in performance of the rank 2 predictor over the rank 1 predictor for short/middle distances over long distances.
Our results may have implications for existing hypotheses and findings in sports science on the differences in physiological determinants of long and short distance running respectively.
These include differences in the muscle fibre types contributing to performance (type I vs. type II) \cite{saltin1977fibre, hoppeler1985endurance}, whether the race length demands energy primarily from aerobic or anaerobic metabolism \cite{bosquet2002methods, faude2009lactate},
which energy systems are mobilized (glycolysis vs. lipolysis) \cite{brooks1994balance, venables2005determinants} and whether the race terminates before the onset of a \.VO$_2$ slow component \cite{borrani2001v, poole1994vo2}.
We conjecture that the combination of our methodology with experiments will shed further light on these differences.

\item An open question in the physiology of aging is whether power or endurance capabilities diminish faster with age. Our analysis provides cross-sectional evidence that {\bf training standard decreases with age, and specialization shifts away from endurance}. This confirms observations of Rittweger et al.~\cite{rittweger2009sprint} on masters world-record data.
There are multiple possible explanations for this, for example longitudinal changes in specialization, or selection bias due to older athletes preferring longer distances; our model renders these hypotheses amenable to quantitative validation.

\item We find that there are a number of {\bf high-standard athletes who attempt distances different from their inferred best distance}; most notably a cluster of young athletes (< 25 ys) who run short distances, and a cluster of older athletes (>40 y) who run long distances, but who we predict would perform better on longer resp.~shorter distances.
Moreover, the third component of our model implies the existence of  {\bf athletes with very strong specialization in their best event}; there are indeed high profile examples of such athletes, such as Zersenay Tadese, who holds the half-marathon world best performance (58:23) but has as yet to produce a marathon performance even close to this in quality (best performance, 2:10:41).
\end{itemize}

We also anticipate that our framework will prove fruitful in {\bf equipping the practioner with new methods for prediction and quantification}:

\begin{itemize}
  \itemsep0em
  \item Individual predictions are crucial in {\bf race planning}---especially for predicting a target performance for events such as the Marathon for which months of preparation are needed;
 the ability to accurately select a realistic target speed will make the difference between an athlete achieving a personal best performance and dropping out of the race from exhaustion.
  \item Predictions and the three-number-summary yield a concise description of the runner's specialization and training state and are thus of immediate use in {\bf training assessment and planning}, for example in determining the potential effect of a training scheme or finding the optimal event(s) for which to train.
  \item The presented framework allows for the derivation of novel and more accurate {\bf scoring schemes} including scoring tables for any type of population.
  \item Predictions for elite athletes allow for a more precise {\bf estimation of quotas and betting risk}. For example, we predict that a fair race between Mo Farah and Usain Bolt is over 492m (374-594m with 95\% prob),
  Chris Lemaitre and Adam Gemili have the calibre to run 43.5 ($\pm1.3$) and 43.2 ($\pm1.3$) resp.~seconds over 400m and Kenenisa Bekele is capable at his best of a 2:00:36  marathon ($\pm3.6$ mins).
 \end{itemize}

We further conjecture that the physiological laws we have validated for running will be immediately transferable to any sport where a power law has been observed on the collective level, such as swimming, cycling, and horse racing.

\section*{Acknowledgments}

We thank Ryota Tomioka for providing us with his code for matrix completion via nuclear norm minimization~\cite{tomioka2010extension}, and for advice on its use. We thank Louis Theran for advice regarding the implementation of local matrix completion in higher ranks.

We thank Klaus-Robert M\"uller for helpful comments on earlier versions of the manuscript.

DAJB was supported by a grant from the German Research Foundation, research training group GRK 1589/1 ``Sensory Computation in Neural Systems''.

FK was partially supported by Mathematisches Forschungsinstitut Oberwolfach (MFO). This research was partially carried out at MFO with the support of FK's Oberwolfach Leibniz Fellowship.

\section*{Author contributions}
DAJB conceived the application LMC to athletic performance prediction, and acquired the data.
DAJB and FJK jointly contributed to methodology and the working form of the LMC prediction algorithm. FJK conceived the LMC algorithm in higher ranks; DAJB designed its concrete implementation and adapted the LMC algorithm for the performance prediction problem. DAJB and FJK jointly designed the experiments and analyses. DAJB carried out the experiments and analyses. DAJB and FJK jointly wrote the paper and are jointly responsible for presentation, discussion and interpretation.

\newpage
\section*{Methods}

The following provides a guideline for reproducing the results.
Raw and pre-processed data in MATLAB and CSV formats is available upon request, subject to approval by British Athletics.
Complete and documented source code of algorithms and analyses can be obtained from [download link will be provided here after acceptance of the manuscript].

\subsection*{Data Source}
The basis for our analyses is the online database {\tt www.thepowerof10.info}, which catalogues British individuals' performances achieved in officially ratified athletics competitions since 1954, including Olympic athletic events (field and non-field events), non-Olympic athletic events,  cross country events and road races of all distances.

With permission of British Athletics, we obtained an excerpt of the database by automated querying of the freely accessible parts of {\tt www.thepowerof10.info}, restricted to ten types of running events: 100m, 200m, 400m, 800m, 1500m, the Mile, 5000m (track and road races), 10000m (track and road races), Half-Marathon and Marathon. Other types of running events were available but excluded from the present analyses; the reasons for exclusion were a smaller total of attempts (e.g. 3000m), a different population of athletes (e.g. 3000m is mainly attempted by younger athletes), and varying conditions (steeplechase/ hurdles and cross-country races).

The data set consists of two tables: \texttt{athletes.csv}, containing records of individual athletes, with fields: athlete ID, gender, date of birth; and \texttt{events.csv}, containing records of individual attempts on running events until August 3, 2013, with fields: athlete ID, event type, date of the attempt, and performance in seconds.

The data set is available upon request, subject to approval by British Athletics.

\subsection*{Data Cleaning}

Our excerpt of the database contains (after error and duplication removal) records of 164,746 individuals of both genders, ranging from the amateur to the elite, young to old, and a total of 1,410,789 individual performances for 10 different types of events
(see previous section).

Gender is available for all athletes in the database  (101,775 male, 62,971 female). The dates of birth of 114,168 athletes are missing (recorded as January 1, 1900 in \texttt{athletes.csv} due to particulars of the automated querying); the date of birth of six athletes is set to missing due to an recorded age at recorded attempts of eight years or less.

For the above athletes, a total of 1,410,789 attempts are recorded: 192,947 over 100m, 194,107 over 200m, 109,430 over 400m, 239,666 over 800m, 176,284 over 1500m, 6,590 at the Mile distance, 96,793 over 5000m (the track and road races), 161,504 over 10000m (on the track and road races), 140,446 for the Half-Marathon and 93,033 for the Marathon. Dates of the attempt are set to missing for 225 of the attempts that record January 1, 1901, and one of the attempts that records August 20, 2038. A number of 44 events is removed from the working data set whose reported performances are better than the official world records of their time, or extremely slow, leaving a total of 1,407,432 recorded attempts in the cleaned data set.

\subsection*{Data Preprocessing}

The events and athletes data sets are collated into $(10\times 164,746)$-tables/matrices of performances, where
the $10$ columns correspond to events and the $164,746$ rows to individual athletes. Rows are indexed increasingly by athlete ID, columns by the type of event. Each entry of the table/matrix contains one performance (in seconds) of the athlete by which the row is indexed, at the event by which the column is indexed, or a missing value. If the entry contains a performance, the date of that performance is stored as meta-information.

We consider two different modes of collation, yielding one table/matrix of performances of size $(10\times 164,746)$ each.

In the first mode, which in Tables~\ref{tab:compare_methods} ff.~is referenced as {\bf ``best''}, one proceeds as follows. First, for each individual athlete, one finds the best event of each individual, measured by population percentile. Then, for each type of event which was attempted by that athlete within a year before that best event, the best performance for that type of event is entered into the table. If a certain event was not attempted in this period, it is recorded as missing.

For the second mode of collation, which in Tables~\ref{tab:compare_methods} ff.~is referenced as {\bf ``random''}, one proceeds as follows. First, for each individual athlete, a calendar year is uniformly randomly selected among the calendar years in which that athlete has attempted at least one event. Then, for each type of event which was attempted by that athlete within the selected calendar year, the best performance for that type of event is entered into the table. If a certain event was not attempted in the selected calendar year, it is recorded as missing.

The first collation mode ensures that the data is of high quality: athletes are close to optimal fitness, since
their best performance was achieved in this time period. Moreover, since fitness was at a high level,
it is plausible that the number of injuries incurred was low; indeed it can be observed that the number of attempts per event is higher in this period, effectively decreasing the influence of noise and the chance that outliers are present after collation.

The second collation mode is used to check whether and, if so how strongly, the results depend on the athletes being close to optimal fitness.

In both cases choosing a narrow time frame ensures that performances are relevant to one another for prediction.

\subsection*{Athlete-Specific Summary Statistics}
For each given athlete, several summaries are computed based on the collated matrix.

Performance {\bf percentiles} are computed for each event which an athlete attempts in relation to the other athletes' performances on the same event. These column-wise event-specific percentiles, yield a percentile matrix with the same filling pattern (pattern of missing entries) as the collated matrix.

The {\bf preferred distance} for a given athlete is the geometric mean of the attempted events' distances. That is, if $s_1,\dots, s_m$ are the distances for the events which the athlete has attempted, then $\tilde{s} = (s_1\cdot s_2\cdot\ldots\cdot s_m)^{1/m}$ is the preferred distance.

The {\bf training standard} for a given athlete is the mean of all performance percentiles in the corresponding row.

The {\bf no.~events} for a given athlete is the number of events attempted by an athlete in the year of the data considered ({\bf best} or {\bf random}).

Note that the percentiles yield a mostly physiological description; the preferred distance is a behavioural summary since it describes the type of events the athlete attempts. The training standard combines both physiological and behavioural characteristics.

Percentiles, preferred distance, and training standard depend on the collated matrix. At any point when rows of the collated matrix are removed, future references to those statistics refer to and are computed for the matrix where those have been removed; this affects the percentiles and therefore the training standard which is always relative to the athletes in the collated matrix.

\subsection*{Outlier Removal}

Outliers are removed from the data in both collated matrices. An outlier score for each athlete/row is obtained as the difference of maximum and minimum of all performance percentile of the athlete. The five percent rows/athletes with the highest outlier score are removed from the matrix.

\subsection*{Prediction: Evaluation and Validation}

Prediction accuracy is evaluated on row-sub-samples of the collated matrices, defined by (a) a potential subgroup, e.g., given by age or gender, (b) degrees-of-freedom constraints in the prediction methods that require a certain amount of entries per row, and (c) a certain performance percentiles of athletes.

The row-sub-samples referred to in the main text and in Tables~\ref{tab:compare_methods} ff.~are obtained by (a) retaining all rows/athletes in the subgroup specified by gender, or age in the best event, (b) retaining all rows/athletes with at least {\bf no.~events} or more entries non-missing, and discarding all rows/athletes with strictly less than {\bf no.~events} entries non-missing, then (c) retaining all athletes in a certain percentile range. The percentiles referred to in (c) are computed as follows: first, for each column, in the data retained after step (b), percentiles are computed. Then, for each row/athlete, the best of these percentiles is selected as the score over which the overall percentiles are taken.

The accuracy of prediction is measured empirically in terms of out-of-sample root mean squared error (RMSE) and mean absolute error (MAE), with RMSE, MAE, and standard deviations estimated from the empirical sample of residuals obtained in 1000 iterations of leave-one-out validation.

Given the row-sub-sample matrix obtained from (a), (b), (c), prediction and thus leave-one-out validation is done in two ways:
(i) predicting the left-out entry from potentially all remaining entries. In this scenario, the prediction method may have access to the performance of the athlete in question which lie in the future of the event to be predicted, though only performances of other events are available;
(ii) predicting the left-out entry from all remaining entries of other athletes, but only from those events of the athlete in question that lie in the past of the event to be predicted. In this task, temporal causality is preserved on the level of the single athlete for whom prediction is done; though information about other athletes' results that lie in the future of the event to be predicted may be used.

The third option (iii) where predictions are made only from past events has not been studied due to the size of the data set which makes collation of the data set for every single prediction per method and group a computationally extensive task, and due to the potential group-wise sampling bias which would be introduced, skewing the measures of prediction-quality---the population of athletes on the older attempts is different in many respects from the more recent attempts.
We further argue that in the absence of such technical issues, evaluation as in (ii) would be equivalent to (iii); since the performances of two randomly picked athletes, no matter how they are related temporally, can in our opinion be modelled as statistically independent; positing the contrary would be equivalent to postulating that any given athlete's performance is very likely to be directly influenced by a large number of other athlete's performance history, which is an assumption that appears to us to be scientifically implausible.
Given the above, due to equivalence of (ii) and (iii), and the issues occurring in (iii) exclusively, we can conclude that (ii) is preferrable over (iii) from a scientific and statistical viewpoint.

\subsection*{Prediction: Target Outcomes}

The principal target outcome for the prediction is ``performance'', which we present to the prediction methods in three distinct parameterisations. This corresponds to passing not the raw performance matrices obtained in the section ``Data Pre-processing'' to the prediction methods, but re-parameterized variants where the non-missing entries undergo a univariate variable transform. The three parameterizations of performance considered in our experiments are the following:\\
(a) {\bf normalized}: performance as the time in which the given athlete (row) completes the event in question (column), divided by the average time in which the event in question (column) is completed in the sub-sample;\\
(b) {\bf log-time}: performance as the natural logarithm of time in seconds in which the given athlete (row) completes the event in question (column);\\
(c) {\bf speed}: performance as the average speed in meters per second, with which the given athlete (row) completes the event in question (column).\\
The words in italics indicate which parameterisation is referred to in Table~\ref{tab:compare_methods}. The error measures, RMSE and MAE, are evaluated in the same parameterisation in which prediction is performed. We do not evaluate performance directly in un-normalized time units, as in this representation performances between 100m and the Marathon span 4 orders of magnitude (base-10), which would skew the measures of goodness heavily towards accuracy over the Marathon.

Unless stated otherwise, predictions are made in the same parameterisation on which the models are learnt.

\subsection*{Prediction: Models and Algorithms}

In the experiments, a variety of prediction methods are used to perform prediction from the performance data, given as described in ``Prediction: Target Outcomes'', evaluated by the measures as described in the section ``Prediction: Evaluation and Validation''.

In the code available for download, each method is encapsulated as a routine which predicts a missing entry when given the (training entries in the) performance matrix. The methods can be roughly divided in four classes: (1) naive baselines, (2) representatives of the state-of-the-art in prediction of running performance, (3) representatives of the state-of-the-art in matrix completion, and (4) our proposed method and its variants.

The naive baselines are:\\
(1.a) {\bf mean}: predicting the the mean over all performances for the same event.
(1.b) {\bf $k$-NN}: $k$-nearest neighbours prediction. The parameter $k$ is obtained as the minimizer of out-of-sample RMSE on five groups of 50 randomly chosen validation data points from the training set, from among $k=1, k=5$, and $k=20$.

The representatives of the state-of-the-art in predicting running performance are:

(2.a) {\bf Riegel}: The Riegel power law formula with exponent 1.06.
(2.b) {\bf power-law}: A power-law predictor, as per the Riegel formula, but with the exponent estimated from the data. The exponent is the same for all athletes and estimated as the minimizer of the residual sum of squares.
(2.c) {\bf ind.power-law}: A power-law predictor, as per the Riegel formula, but with the exponent estimated from the data. The exponent may be different for each athlete and is estimated as the minimizer of the residual sum of squares.
(2.d) {\bf Purdy}: Prediction by calculation of equivalent performances using the Purdy points scheme~\cite{purdy1974computer}. Purdy points are calculated by using the measurements given by the Portugese scoring tables which estimate the maximum velocity for a given distance in a straight line, and adjust for the cost of traversing curves and the time required to reach race velocity.
The performance with the same number of points as the predicting event is imputed.

The representatives of the state-of-the-art in matrix completion are:

(3.a) {\bf EM}: Expectation maximization algorithm assuming a multivariate Gaussian model for the rows of the performance matrix in log-time parameterisation. Missing entries are initialized by the mean of each column. The updates are terminated when the percent increase in log-likelihood is less than 0.1\%. For a review of the EM-algorithm see~\cite{bishop2006pattern}.
(3.b) {\bf Nuclear Norm}:  Matrix completion via nuclear norm minimization~\cite{candes2009exact, tomioka2010extension}, in the regularized version and implementation by~\cite{tomioka2010extension}.

The variants of our proposed method are as follows:

(4.a-d) {\bf LMC rank $r$}: local matrix completion for the low-rank model, with rank $r = 1,2,3,4$. (4.a) is LMC rank 1, (4.b) is LMC rank 2, and so on.

Our algorithm follows the local/entry-wise matrix completion paradigm in~\cite{kiraly15matrixcompletion}. It extends the rank $1$ local matrix completion method described in~\cite{KirThe13RankOneEst} to arbitrary ranks.

Our implementation uses: determinants of size $(r+1 \times r+1)$ as the only circuits; the weighted variance minimization principle in~\cite{KirThe13RankOneEst}; the linear approximation for the circuit variance outlined in the appendix of~\cite{blythe2014algebraic}; modelling circuits as independent for the co-variance approximation.

We further restrict to circuits supported on the event to be predicted and the $r$ log-distance closest events.

For the convenience of the reader, we describe  the exact way in which the local matrix completion principle is instantiated, in the section ``Prediction: Local Matrix Completion'' below

In the supplementary experiments we also investigate two aggregate predictors to study the potential benefit of using other lengths for prediction:

(5.a) {\bf bagged power law}: bagging the power law predictor with estimated coefficient (2.b) by a weighted average of predictions obtained from different events. The weighting procedure is described below.
(5.b) {\bf bagged LMC rank 2}: estimation by LMC rank 2 where determinants can be supported at any three events, not only on the closest ones (as in line 1 of Algorithm~\ref{alg:LMC} below). The final, bagged predictor is obtained as a weighted average of LMC rank 2 running on different triples of events. The weighting procedure is described below.

The averaging weights for (5.a) and (5.b) are both obtained from the Gaussian radial basis function kernel $\exp \left( \gamma \Delta \Delta^\top\right)$, where $\Delta = \text{log}(\mathbf{s}_{p})-\text{log}(s_{p'})$
and $\mathbf{s}_{p}$ is the vector of predicting distances and $s_{p'}$ is the predicted distance. The kernel width $\gamma$ is a parameter of the bagging. As $\gamma$ approaches 0, aggregation approaches averaging and thus the ``standard'' bagging predictor. As $\gamma$ approaches $-\infty$, the aggregate prediction approaches the non-bagged variants (2.b) and (4.b).

\subsection*{Prediction: Local Matrix Completion}
The LMC algorithm we use is an instance of Algorithm~5 in~\cite{kiraly15matrixcompletion}, where, as detailed in the last section, the circuits are all determinants, and the averaging function is the weighted mean which minimizes variance, in first order approximation, following the strategy outlined in~\cite{KirThe13RankOneEst} and~\cite{blythe2014algebraic}.

The LMC rank $r$ algorithm is described below in pseudo-code. For readability, we use bracket notation $M[i,j]$ (as in R or MATLAB) instead of the usual subscript notation $M_{ij}$ for sub-setting matrices. The notation $M[:,(i_1,i_2,\dots, i_r)]$ corresponds to the sub-matrix of $m$ with columns $i_1,\dots, i_r$. The notation $M[k,:]$ stands for the whole $k$-th row. Also note that the row and column removals in Algorithm~\ref{alg:LMC} are only temporary for the purpose of computation, within the boundaries of the algorithm, and do not affect the original collated matrix.

\begin{algorithm}[ht]
\caption{ -- Local Matrix Completion in Rank $r$.\newline
\textit{Input:} An athlete $a$, an event $s^*$, the collated data matrix of performances $M$. \newline
\textit{Output:} An estimate/denoising for the entry $M[a,s^*]$ \label{alg:LMC}}
\begin{algorithmic}[1]
    \State Determine distinct events $s_1,\dots, s_r\neq s^*$ which are log-closest to $s^*$, i.e., minimize $\sum_{i=1}^r( \log s_i-\log s^*)^2$
    \State Restrict $M$ to those events, i.e., $M\gets M[:,(s^*,s_1,\dots, s_r)]$
    \State Let $v$ be the vector containing the indices of rows in $M$ with no missing entry.
    \State $M\gets M[(v,a),:]$, i.e., remove all rows with missing entries from $M$, except $a$.
    \For{ $i = 1$ to $400$ }
    \State Uniformly randomly sample distinct athletes $a_1,\dots,a_r\neq a$ among the rows of $M$.
    \State Solve the circuit equation $\det M[(a,a_1,\dots, a_r),(s^*,s_1,\dots, s_r)] = 0$ for $s^*$, obtaining a number $m_i$.
    \State Let $A_0,A_1\gets M[(a,a_1,\dots, a_r),(s^*,s_1,\dots, s_r)]$.
    \State Assign $A_0[a,s^*]\gets 0$, and $A_1[a,s^*]\gets 1$.
    \State Compute $\sigma_i\gets\frac{1}{|\det A_0 + \det A_1|}+\frac{|\det A_0|}{(\det A_0 - \det A_1)^2}$
    \State Assign the weight $w_i\gets \sigma_i^{-2}$
    \EndFor
    \State Compute $m^*\gets \left(\sum_{i=1}^{400}w_im_i\right)\cdot\left(\sum_{i=1}^{400} w_i\right)^{-1}$
    \State Return $m^*$ as the estimated performance.
\end{algorithmic}
\end{algorithm}

The bagged variant of LMC in rank $r$ repeatedly runs LMC rank $r$ with choices of events different from the log-closest, weighting the results obtained from different choices of $s_1,\dots, s_r$. The weights are obtained from $5$-fold cross-validation on the training sample.

\subsection*{Obtaining the Low-Rank Components and Coefficients}
We obtain three low-rank components
$f_1,\dots, f_3$ and corresponding coefficients $\lambda_{1},\dots, \lambda_{3}$ for each athlete by considering the data in log-time coordinates. Each component $f_i$ is a vector of length 10, with entries corresponding to events. Each coefficient is a scalar, potentially different per athlete.

To obtain the components and coefficients, we consider the data matrix for the specific target outcome, sub-sampled to contain the athletes who have attempted four or more events and the top 25\% percentiles, as described in ``Prediction: Evaluation and Validation''. In this data matrix, all missing values are imputed using the rank 3 local matrix completion algorithm, as described in (4.c) of ``Prediction: Models and Algorithms'', to obtain a complete data matrix $M$. For this matrix, the singular value decomposition $M=USV^\top$ is computed, see~\cite{golub1970singular}.

We take the components $f_2,f_3$ to be the the $2$-th and $3$-rd right singular vectors, which are the $2$-nd and $3$-rd column of $V$. The component $f_1$ is a re-scaled version of the $1$-st column $v$ of $V$, such that $f_1(s) \approx \log s$, where the natural logarithm is taken. More precisely, $f_1 := \alpha v$, where the re-scaling factor $\alpha$ is the minimizer of the sum of squared residuals of $\alpha f_1(s) - \log (s)$ over $s$ being the ten event distanes.

The three-number-summary referenced in the main corpus of the manuscript is obtained as follows: for the $k$-th athlete we obtain from the left singular vector the entries $U_{kj}$. The second and third score of the three-number-summary are obtained as $\lambda_2 = U_{k2}$ and $\lambda_3= U_{k3}$. The individual exponent is $\lambda_1 = \alpha^{-1}\cdot U_{j1}$.

The singular value decomposition has the property that the $f_i$ and $\lambda_j$ are guaranteed to be least-squares estimators for the components and the coefficients in a projection sense.

\subsection*{Computation of standard error and significance}

Standard errors for the singular vectors (components of the model of Equation~\ref{eq:model}) are computed via independent bootstrap sub-sampling on the rows of the data set (athletes).

Standard errors for prediction accuracies are obtained by bootstrapping of the predicted performances (1000 per experiment).
A method is considered to perform significantly better than another when error regions at the 95\% confidence level (= mean over repetitions $\pm$ 1.96 standard errors) do not intersect.

\subsection*{Predictions and three-number-summary for elite athletes}

Performance predictions and three-number-summaries for the selected elite athletes in Table~\ref{tab:elite_scores} and Figure~\ref{fig:scatter} are obtained from their personal best times. The relative standard error of the predicted performances is estimated to be the same as the relative RMSE of predicting time, as reported in Table~\ref{tab:compare_methods}.

\subsection*{Calculating a fair race}

Here we describe the procedure for calculating a fair racing distance with error bars between two athletes: athlete 1 and athlete 2. We first calculate predictions
for all events. Provided that athlete 1 is quicker on some events and athlete 2 is quicker on others, then calculating a fair race is feasible.
If athlete 1 is quicker on shorter events then athlete 2 is typically quicker on all longer events beyond a certain distance. In that case, we can find the shortest race $s_i$ whereby athlete 2 is predicted to be
quicker; then a fair race lies between $s_i$ and $s_{i-1}$. The performance curves in log-time vs. log-distance of both athletes will be locally approximately linear.
We thus interpolate the performance curves between $\text{log}(s_i)$ and $\text{log}(s_{i-1})$---the crossing point gives the position of a fair race in log-coordinates.
We obtain confidence intervals by repeating this procedure after sampling data points around the estimated performances with standard deviation equal to the RMSE (see Table~\ref{tab:compare_methods}) on the top 25\% of athletes in log-time.

\newpage
\section*{Supplementary Analyses}

This appendix contains a series of additional experiments supplementing those in the main corpus. It contains the following findings:\\

{\bf (S.I) Validation of the LMC prediction framework.}\\
{\bf (S.I.a) Evaluation in terms of MAE.} The results in terms of MAE are qualitatively similar to those in RMSE; smaller MAEs indicate the presence of outliers.\\
{\bf (S.I.b) Evaluation in terms of time prediction.} The results are qualitatively similar to measuring prediction accuracy in RMSE and MAE of log-time. LMC rank 2 has an average error of approximately 2\% when predicting the top 25\% of male athletes.\\
{\bf (S.I.c) Prediction for individual events.} LMC outperforms the other predictors on each type of event. The benefit of higher rank is greatest for middle distances.\\
{\bf (S.I.d) Stability w.r.t.~the unit measuring performance.} LMC performs equally well in predicting (performance in time units) when performances are presented in log-time or time normalized by event average. Speed is worse when the rank 2 predictor is used.\\
{\bf (S.I.e) Stability w.r.t.~the events used in prediction.} LMC performs equally well when predicting from the closest-distance events and when using a bagged version which uses all observed events for prediction.\\
{\bf (S.I.f) Stability w.r.t.~the event predicted.} LMC performs well both when the predicted event is close to those observed and when the predicted event is further from those observed, in terms of event distance.\\
{\bf (S.I.g) Temporal independence of performances.} There are no differences between predictions made only from past events and predictions made from all available events (in the training set).\\
{\bf (S.I.h) Run-time comparisons.} LMC is by orders of magnitude the fastest among the matrix completion methods.\\

{\bf (S.II) Validation of the low-rank model.}\\
{\bf (S.II.a) Synthetic validation.} In a synthetic low-rank model of athletic performance that is a proxy to the real data, the singular components of the model can be correctly recovered by the exact same procedure as on the real data. The generative assumption of low rank is therefore appropriate.\\
{\bf (S.II.b) Universality in sub-groups.} Quality of prediction, the low-rank model, its rank, and the singular components remain mostly unchanged when considering subgroups male/female, older/younger, elite/amateur.\\

{\bf (S.III) Exploration of the low-rank model.}\\
{\bf (S.III.a) Further exploration of the three-number-summary.} The three number summary also correlates with specialization and training standard.\\
{\bf (S.III.b) Preferred distance vs optimal distance.} Most but not all athletes prefer to attend the event at which they are predicted to perform best. A notable number of younger athletes prefer distances shorter than optimal, and some older athletes prefer distances longer than optimal.\\

{\bf (S.IV) Pivoting and phase transitions.} The pivoting phenomenon in Figure~\ref{fig:illustration}, right panel, is found in the data for any three close-by distances up to the Mile, with anti-correlation between the shorter and the longer distance. Above 5000m, a change in the shorter of the three distances positively correlates with a change in the longer distance.\\

\noindent
{\bf (S.I.a) Evaluation in terms of MAE.} Table~\ref{tab:compare_methods_mae} reports on the goodness of prediction methods in terms of MAE. Compared with the RMSE (Table~\ref{tab:compare_methods}, the MAE tend to be smaller than the RMSE, indicating the presence of outliers. The relative prediction-accuracy of methods when compared to each other is qualitatively the same.\\

\noindent
{\bf (S.I.b) Evaluation in terms of time prediction.} Tables~\ref{tab:rrmse_time} and~\ref{tab:rmae_time} report on the prediction accuracy of the methods tested in terms the relative RMSE and MAE of predicting time. Relative measures are chosen to avoid bias towards the longer events. The results are qualitatively and quantitatively very similar to the log-time results in Tables~\ref{tab:compare_methods} and~\ref{tab:compare_methods_mae}; this can be explained that mathematically the RMSE and MAE of a logarithm approximate the relative RMSE and MAE well for small values.\\

\begin{figure}
\begin{center}
$\begin{array}{c}
\includegraphics[width=100mm,clip=true,trim= 0mm 0mm 0mm 0mm]{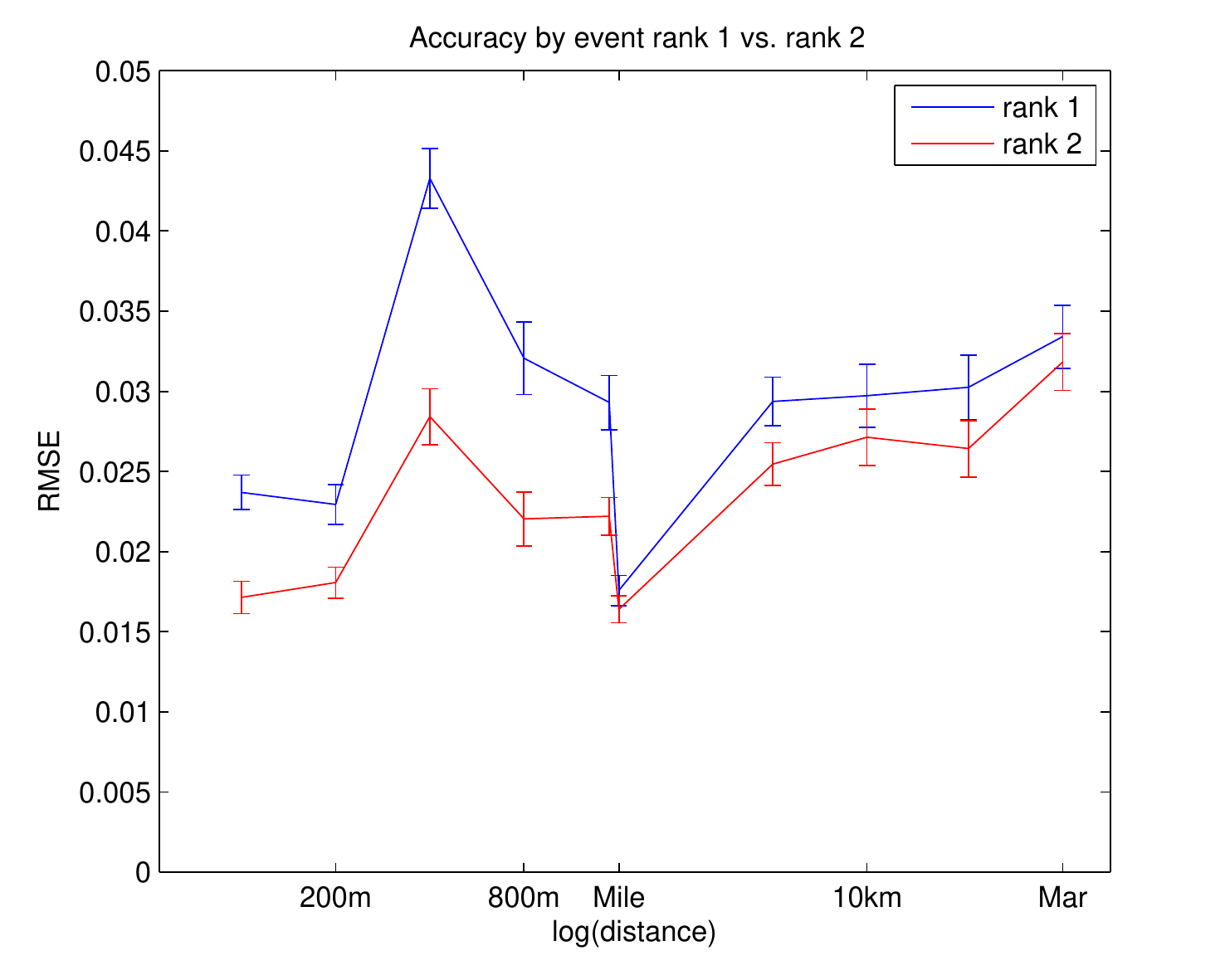}
\end{array}$
\end{center}
\caption{The figure displays the results of prediction by event for the top 25\% of male athletes who attended $\geq3$ events in their year of best performance. For each
event the prediction accuracy of LMC in rank 1 (blue) is compared to prediction accuracy in rank 2 (red).
RMSE is displayed on the $y$-axis against distance on the $x$-axis; the error bars extend two standard deviations of the bootstrapped RMSE either side of the RMSE.
\label{fig:individual_events}}
\end{figure}

\noindent
{\bf (S.I.c) Individual Events.}
Prediction accuracy of LMC rank 1 and rank 2 on the ten different events is displayed in Figure~\ref{fig:individual_events}.
The reported prediction accuracy is out-of-sample RMSE of predicting log-time, on the top 25 percentiles of Male athletes who have attempted 3 or more events, of events in their best year of performance. The reported RMSE for a given event is the mean over 1000 random prediction samples, standard errors are estimated by the bootstrap.\\
The relative improvement of rank 2 over rank 1 tends to be greater for shorter distances below the Mile. This is in accordance with observation (IV.i) which indicates that the individual exponent is the best descriptor for longer events, above the Mile.\\

\noindent
{\bf (S.I.d) Stability w.r.t.~the measure of performance.} In the main experiment, the LMC model is learnt on the same measure of performance (log-time, speed, normalized) which is predicted. We investigate whether the measure of performance on which the model is learnt influences the prediction by learning the LMC model on either measure and comparing all predictions using the log-time measure. Table~\ref{tab:choose_feature} displays prediction accuracy when the model is learnt in any one of the measures of performance. Here we check the effect of calibration in one coordinates system and testing in another. The reported goodness is out-of-sample RMSE of predicting log-time, on the top 25 percentiles of Male athletes who have attempted 3 or more events, of events in their best year of performance. The reported RMSE for a given event is the mean over 1000 random prediction samples, standard errors are estimated by the bootstrap.\\
We find that there is no significant difference in prediction goodness when learning the model in log-time coordinates or normalized time coordinates. Learning the model in speed coordinates leads to a significantly better prediction than log-time or normalized time when LMC rank 1 is applied, but to a worse prediction with LMC rank 2. As overall prediction with LMC rank 2 is better, log-time or normalized time are the preferable units of performance.\\

\noindent
{\bf (S.I.e) Stability w.r.t.~the event predicted.}

\begin{figure}
\begin{center}
$\begin{array}{c}
\includegraphics[width=140mm,clip=true,trim= 20mm 0mm 20mm 0mm]{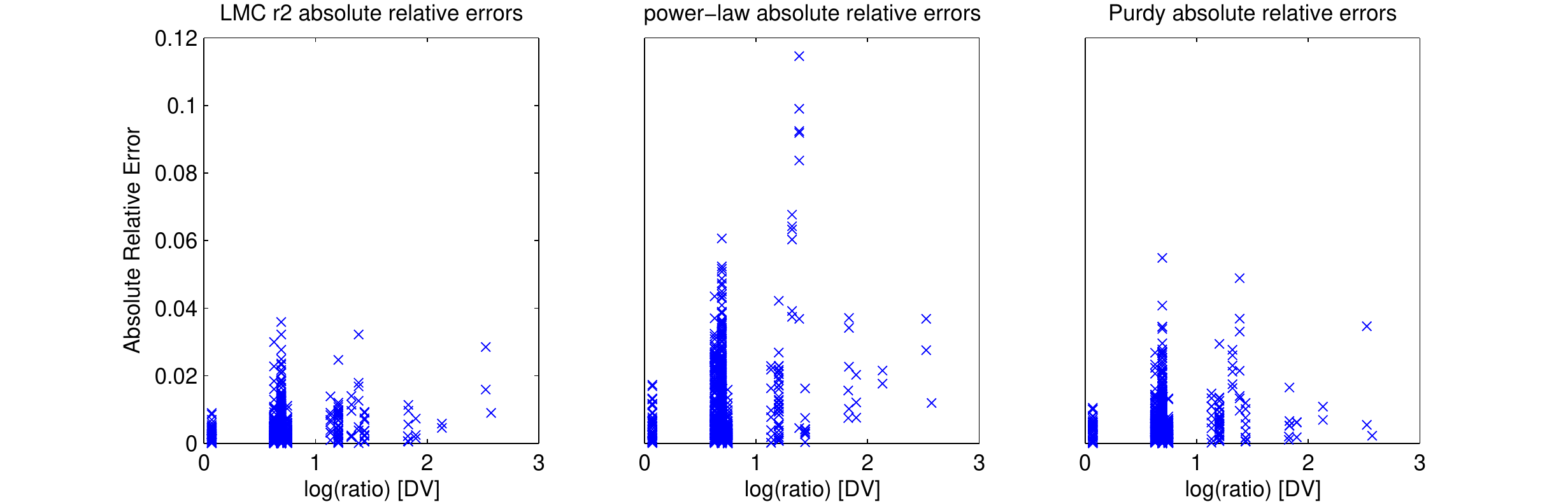}
\end{array}$
\end{center}
\caption{The figure displays the absolute log ratio in distance predicted and predicting distance vs. absolute relative error per athlete.
In each case the log ratio in distance is displayed on the $x$-axis and the absolute errors of single data points of the $y$-axis.
We see that LMC in rank 2 is particularly robust for large ratios in comparison to the power-law and Purdy Points. Data is taken from
the top 25\% of male athletes with {\bf no.~events}$\geq 3$ in the {\bf best }year.
 \label{fig:by_deviation}}
\end{figure}

We consider here the effect of the ratio between the predicted event and the closest predictor. For data of the best 25\% of Males
in the year of best performance ({\bf best}), we compute the log-ratio of the closest predicting distance and the predicted distance for Purdy Points,
the power-law formula and LMC in rank 2.
See Figure~\ref{fig:by_deviation}, where this log ratio is plotted by error. The results show that LMC is far more robust to
error for predicting distances far from the predicted distance.\\

\noindent
{\bf (S.I.f) Stability w.r.t.~the events used in prediction.} We compare whether we can improve prediction by using all events an athlete has attempted, by using one of the aggregate predictors (5.a) bagged power law or (5.b) bagged LMC rank 2. The kernel width $\gamma$ for the aggregate predictors is chosen from $-0.001,-0.01,-0.1,-1,-10$ as the minimizer of out-of-sample RMSE on five groups of 50 randomly chosen validation data points from the training set. The validation setting is the same as in the main prediction experiment.\\
Results are displayed in Table~\ref{tab:learn_weights}. We find that prediction accuracy of (2.b) power law and (5.a) bagged power law is not significantly different, nor is (4.b) LMC rank 2 significantly different from (5.b) bagged LMC rank 2 (both $p>0.05$; Wilcoxon signed-rank on the absolute residuals). Even though the kernel width selected is in the majority of cases $\sigma = -1$ and not $\sigma = -10$, the incorporation of all events does not lead to an improvement in prediction accuracy in our aggregation scheme.
We find there is no significant difference ($p>0.05$; Wilcoxon signed-rank on the absolute errors) between the bagged and vanilla LMC for the top 95\% of runners. This demonstrates
that the relevance of closer events for prediction may be learn from the data. The same holds for the bagged version of the power-law formula. \\

\noindent
{\bf (S.I.g) Temporal independence of performances.}
We check here whether the results are affected by using only temporally prior attempts in predicting an athlete's performance, see section ``Prediction: Evaluation and Validation'' in ``Methods''. To this end, we compute out-of-sample RMSEs when predictions are made only from those events.

Table~\ref{tab:rrmse_causal} reports out-of-sample RMSE of predicting log-time, on the top 25 percentiles of Male athletes who have attempted 3 or more events, of events in their best year of performance. The reported RMSE for a given event is the mean over 1000 random prediction samples, standard errors are estimated by the bootstrap.

 The results are qualitatively similar to those of Table~\ref{tab:compare_methods} where all events are used in prediction.\\

\begin{figure}
\begin{center}
$\begin{array}{c}
\includegraphics[width=60mm,clip=true,trim= 0mm 0mm 0mm 0mm]{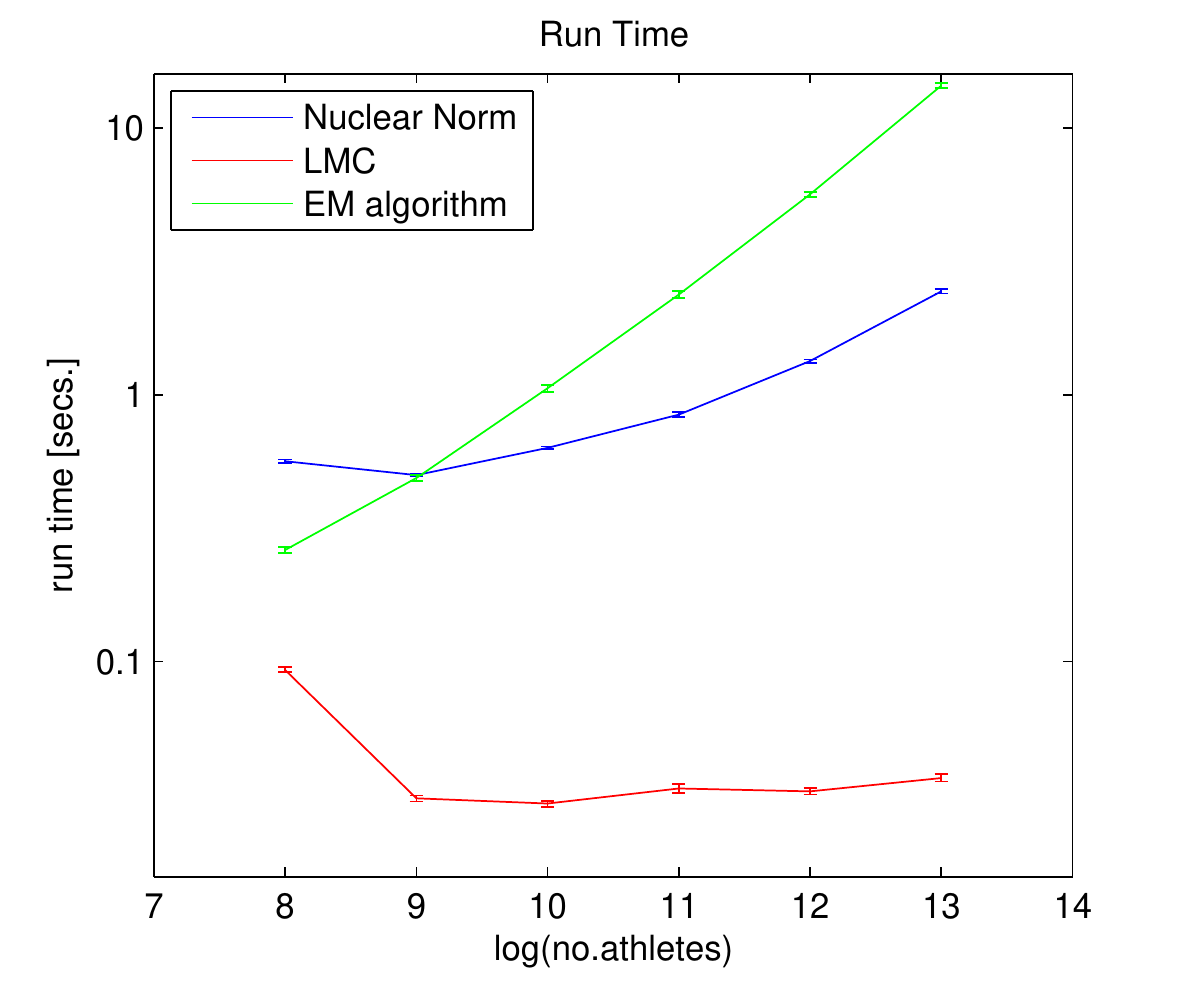}
\end{array}$
\end{center}
\caption{The figure displays mean run-times for the 3 matrix completion algorithms tested in the paper: Nuclear Norm, EM and LMC (rank 3). Run-times (y-axis) are recorded for completing a single entry in a matrix of size indicated by the x-axis. The averages are over 100 repetitions, standard errors are estimated by the bootstrap.
\label{fig:runtime}}
\end{figure}

\noindent
{\bf (S.I.h) Run-time comparisons.} We compare the run-time cost of a single prediction for the three matrix completion methods LMC, nuclear norm minimiziation, and EM. The other (non-matrix completion) methods are fast or depend only negligibly on the matrix size. We measure run time of LMC rank 3 for completion of a single entry for matrices of $2^8,2^9,\dots,2^{13}$ athletes, generated as described in (S.II.a). This is repeated 100 times.
For a
fair comparison, the nuclear norm minimization algorithm is run with a hyper-parameter already pre-selected by cross validation.
The results are displayed in Figure~\ref{fig:runtime}; LMC is faster by orders of magnitude than nuclear norm and EM and is very robust to the
size of the matrix. The reason computation speeds up over the smallest matrix sizes is that $4\times 4$ minors, which are required for rank 3 estimation
are not available, thus the algorithm must attempt all ranks lower than 3 to find sufficiently many minors.
\\

\noindent
{\bf (S.II.a) Synthetic validation.} To validate the assumption of a low-rank generative model, we investigate prediction accuracy and recovery of singular vectors in a synthetic model of athletic performance.

Synthetic data for a given number of athletes is generated as follows:

For each athlete, a three-number summary $(\lambda_1,\lambda_2,\lambda_3)$ is generated independently from a Gaussian distribution with the same mean and variance as the three-number-summaries measured on the real data and with uncorrelated entries.

Matrices of performances are generated from the model
\begin{align}
\text{log}(t)= \lambda_1 f_1(s) + \lambda_2 f_2(s) +  \lambda_3 f_3(s)+ \eta(s) \label{eq:synthetic_model}
\end{align}
where $f_1,f_2,f_3$ are the three components estimated from the real data and $\eta(s)$ is a stationary zero-mean Gaussian white noise process with adjustable variance. We take the components estimated in log-time coordinates from the top 25\% of male athletes who have attempted at least 4 events as the three components of the model. The distances $s$ are the same ten event distances as on the real data. In each experiment the standard deviation of $\eta(s)$

{\bf Accuracy of prediction:} We synthetically generate a matrix of $1000$ athletes according to the model of Equation~\eqref{eq:synthetic_model}, taking
as distances the same distances measured on the real data.
Missing entries are randomized according to two schemes:
(a) 6 (out of 10) uniformly random missing entries per row/athlete.
(b) per row/athlete, four in terms of distance-consecutive entries are non-missing, uniformly at random.

\begin{figure}
\begin{center}
\includegraphics[width = 100mm]{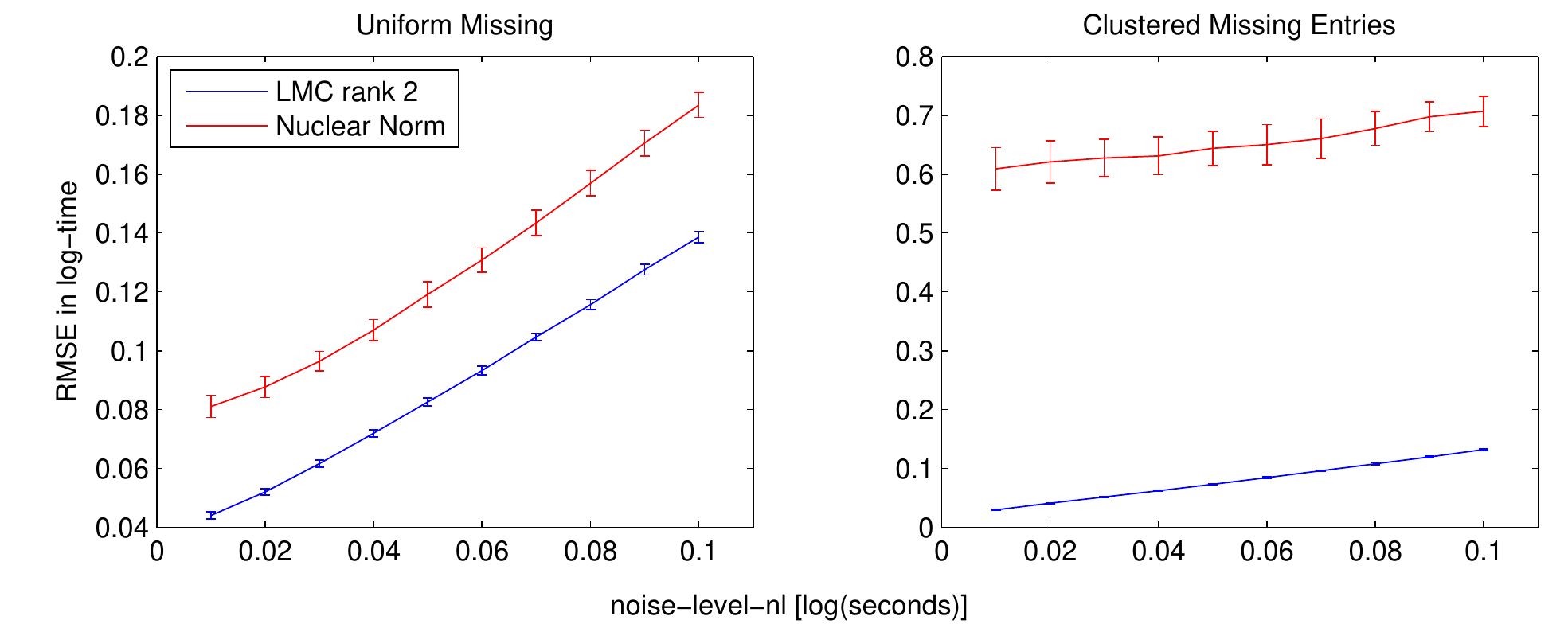}
\end{center}
\caption{LMC and Nuclear Norm prediction accuracy on the synthetic low-rank data. $x$-axis denotes the noise level (standard deviation of additive noise in log-time coordinates); $y$-axis is out-of-sample RMSE predicting log-time. Left: prediction performance when (a) the missing entries in each ros are uniform. Right: prediction performance when (b) the observed entries are consecutive. Error bars are one standard deviation, estimated by the bootstrap.}
\label{fig:toy_accuracy}
\end{figure}

We then apply LMC rank 2 and nuclear norm minimization for prediction. This setup is repeated 100 times for ten different standard deviations of $\eta$ between 0.01 and 0.1. The results are displayed in Figure~\ref{fig:toy_accuracy}.

LMC performance outperforms nuclear norm; LMC performance is also robust to the pattern of missingness, while nuclear norm minimization is negatively affected by clustering in the rows. RMSE of LMC approaches zero with small noise variance, while RMSE of nuclear norm minimization does not.

Comparing the performances with Table~\ref{tab:compare_methods}, an assumption of a noise variance of $\mbox{Std}(\eta) = 0.01$ seems plausible. The performance of nuclear norm on the real data is explained by a mix of the sampling schemes (a) and (b).

{\bf Recovery of model components.} We synthetically generate a matrix which has a size and pattern of observed entries identical to the matrix of top 25\% of male athletes who have attempted at least 4 events in their best year.
We set $\mbox{Std}(\eta) = 0.01$, which was shown to be plausible in the previous section.

We then complete all missing entries of the matrix using LMC rank 3. After this initial step we estimate singular components
using SVD, exactly as on the real data. Confidence intervals are estimated by a bootstrap on the rows with 100 iterations.

\begin{figure}
\begin{center}
\includegraphics[width = 100mm]{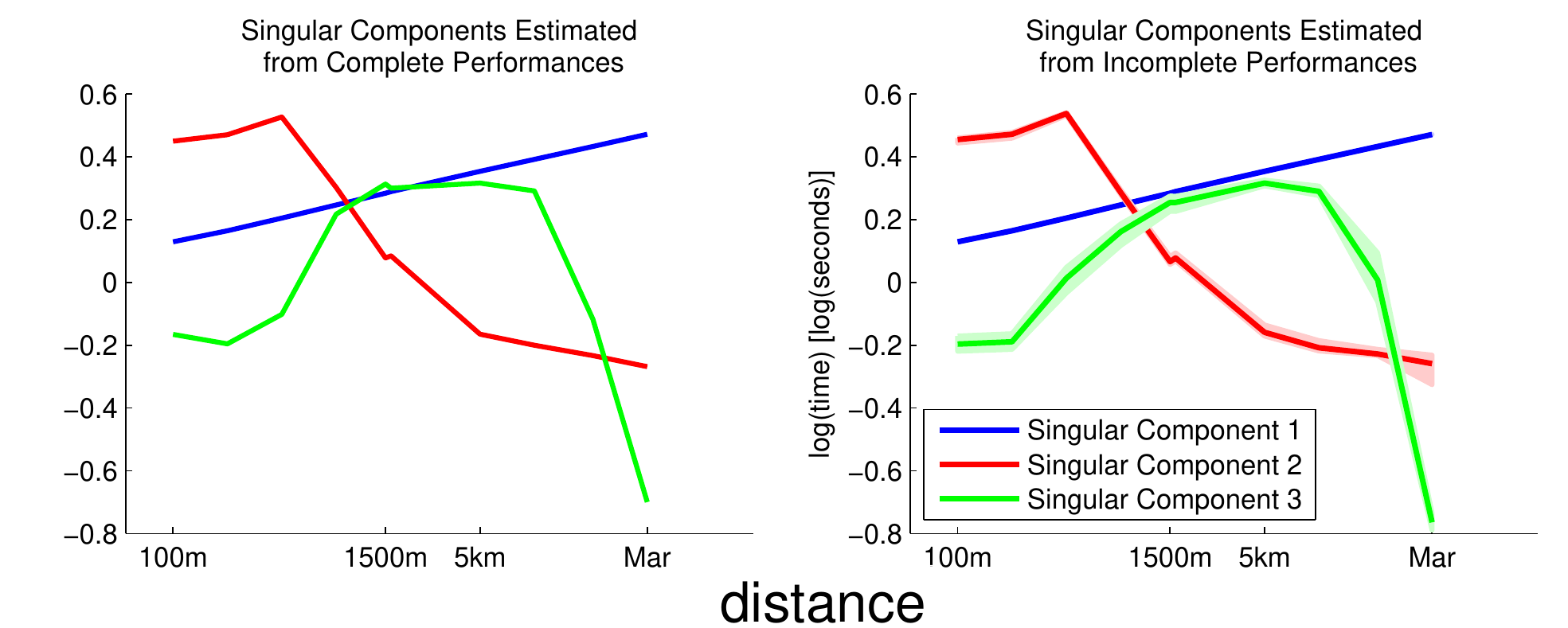}
\end{center}
\caption{Accuracy of singular component estimation with missing data on synthetic model of performance. $x$-axis is distance, $y$-axis is components in log-time.
Left: singular components of data generated according to Equation~\ref{eq:synthetic_model} with all data present. Right: singular components of data generated according to Equation~\ref{eq:synthetic_model} with missing entries
estimated with LMC in rank 3; the observation pattern and number of athletes is identical to the real data. The tubes denote one standard deviation estimated by the bootstrap.}
\label{fig:accuracy_singular_vectors}
\end{figure}

The results are displayed in Figure~\ref{fig:accuracy_singular_vectors}.

One observes that the first two singular components are recovered almost exactly, while the third is a slightly deformed. This is due to the smaller singular value of the third component.
\\

\noindent
{\bf (S.II.b) Universality in sub-groups.} We repeat the methodology for component estimation described above and obtain the three components in the following sub-groups:
female athletes, older athletes (> 30 years), and amateur athletes (25-95 percentile range of training standard). Male athletes were considered in the main corpus. For female and older athletes, we restrict to the top 95\% percentiles of the respective groups for estimation.

\begin{figure}
\begin{center}
$\begin{array}{c c c}
\includegraphics[width=50mm,clip=true,trim= 0mm 0mm 0mm 0mm]{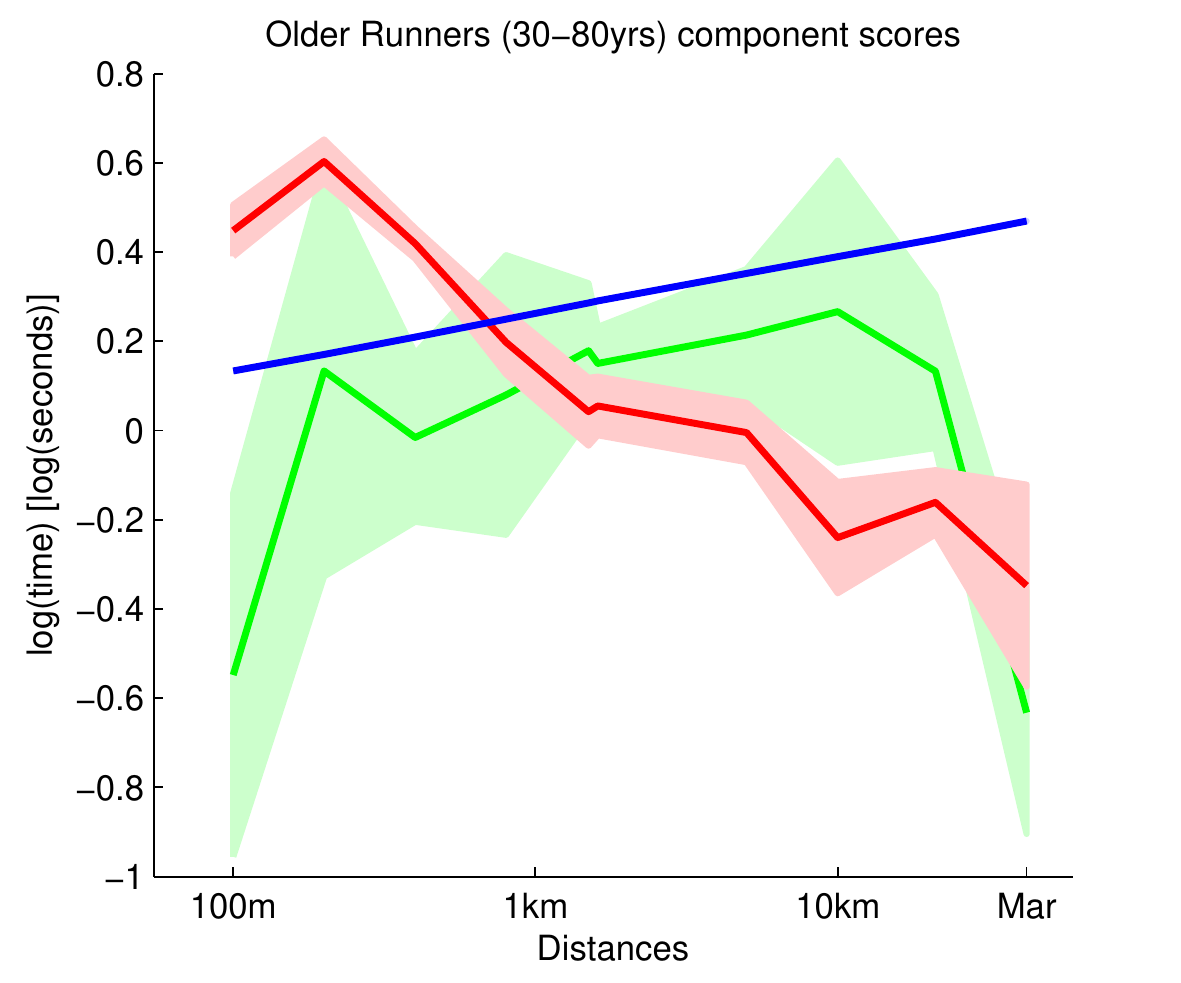}  &
\includegraphics[width=50mm,clip=true,trim= 0mm 0mm 0mm 0mm]{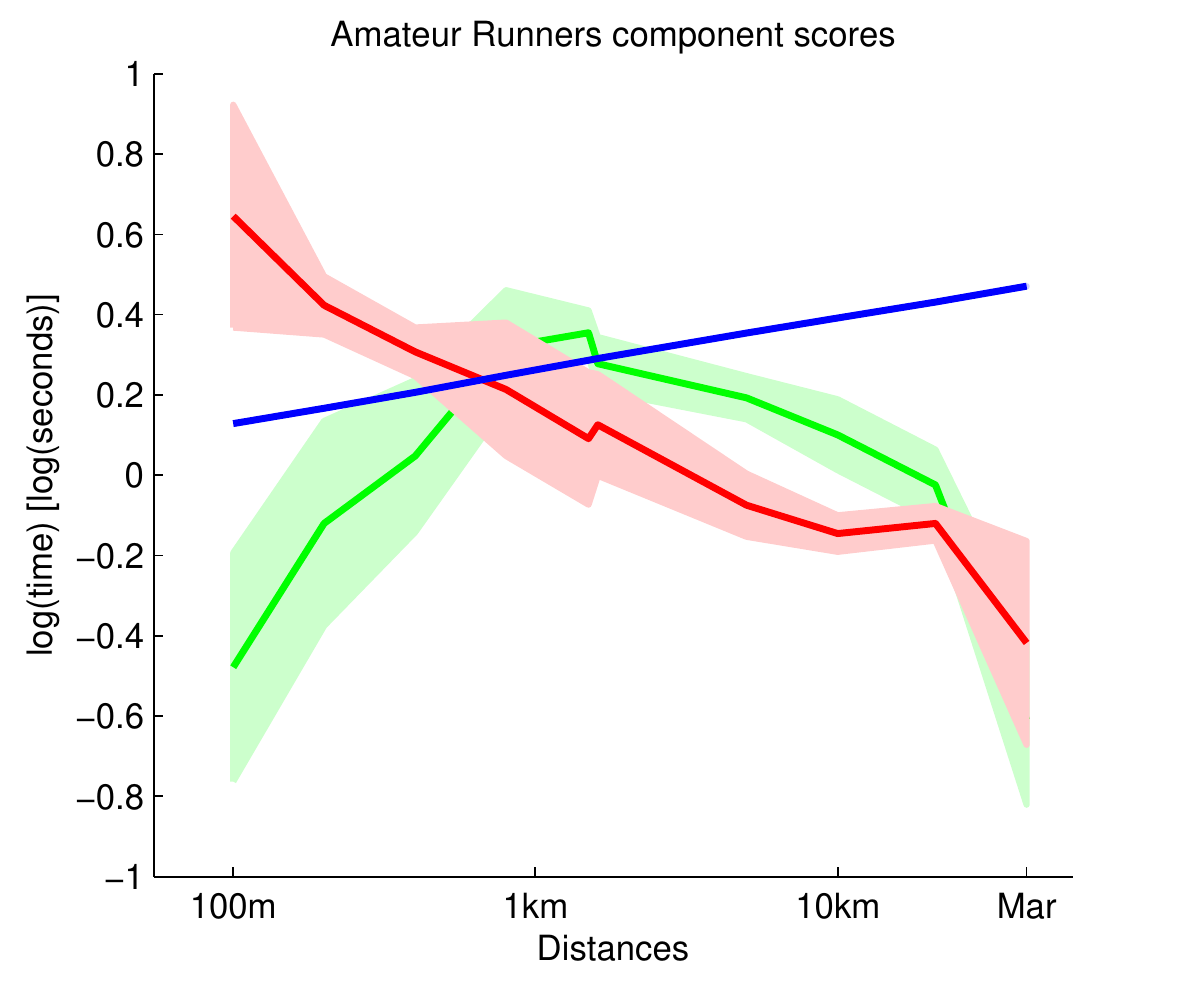}  &
\includegraphics[width=50mm,clip=true,trim= 0mm 0mm 0mm 0mm]{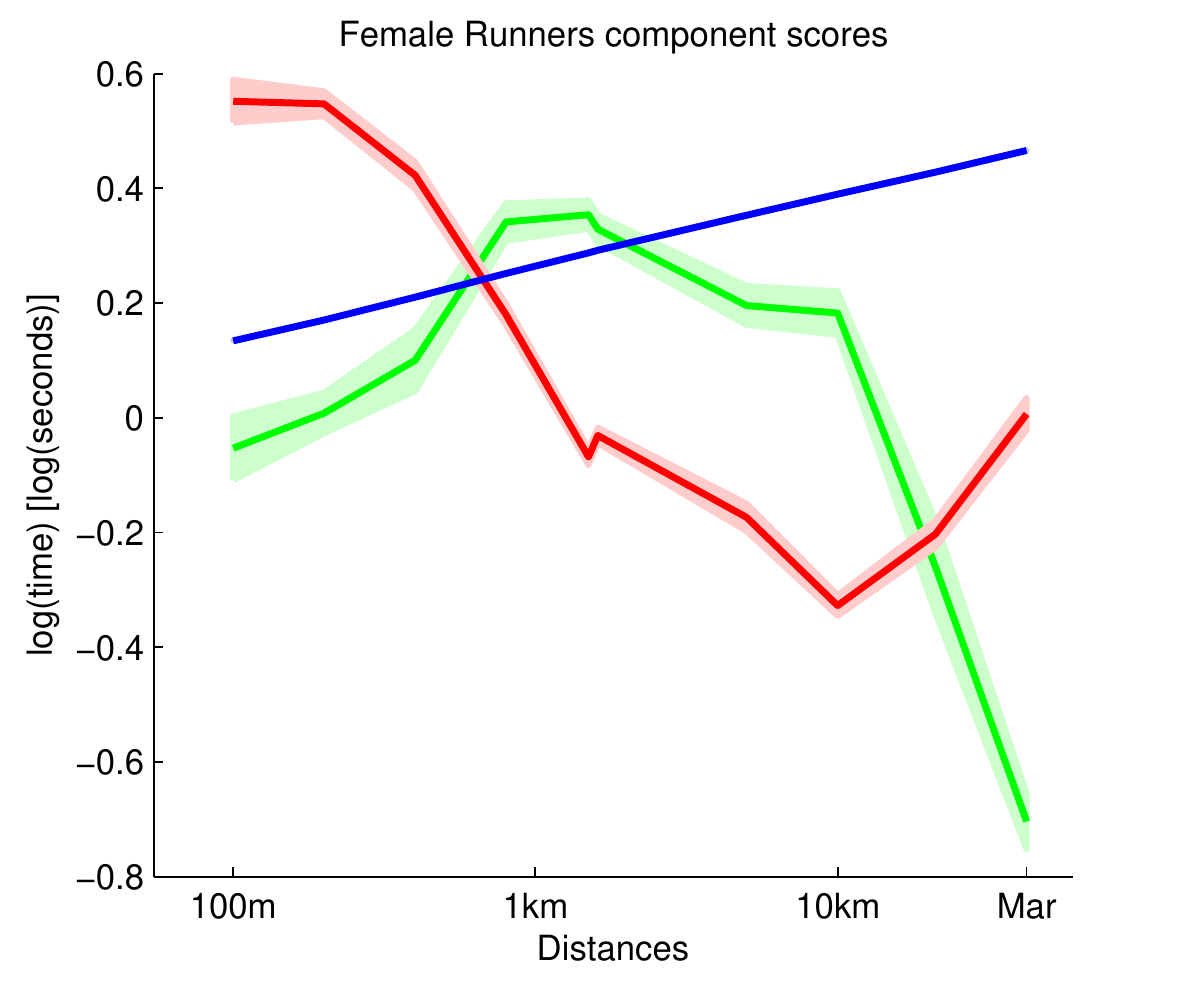}
\end{array}$
\end{center}
\caption{The three components of the low-rank model in subgroups. Left: for older runners. Middle: for amateur runners = best event below 25th percentile. Right: for female runners. Tubes around the components are one standard deviation, estimated by the bootstrap. The components are the analogous components for the subgroups described as computed in the left-hand panel of Figure~\ref{fig:svs}. \label{fig:svs_subgroups}}
\end{figure}

Figure~\ref{fig:svs_subgroups} displays the estimated components of the low-rank model. The individual power law is found to be unchanged in all groups considered. The second and third component vary between the groups but resemble the components for the male athletes.  The empirical variance of the second and third component is higher, which may be explained by a slightly reduced consistency in performance, or a reduction in sample size. Whether there is a genuine difference in form or whether the variation is explained by different three-number-summaries in the subgroups cannot be answered from the dataset considered.

Table~\ref{tab:generalize} displays the prediction results in the three subgroups. Prediction accuracy is similar but slightly worse when compared to the male athletes. Again this may be explained by reduced consistency in the subgroups' performances.\\

\begin{figure}
\begin{center}
$\begin{array}{c}
\includegraphics[width=130mm,clip=true,trim= 0mm 0mm 0mm 0mm]{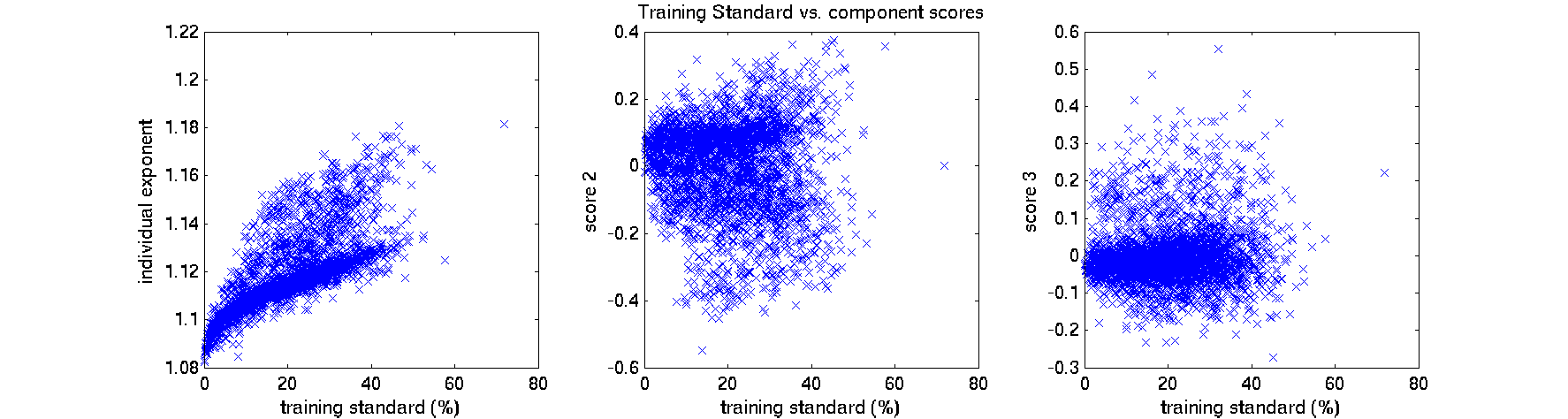}   \\
\includegraphics[width=130mm,clip=true,trim= 0mm 0mm 0mm 0mm]{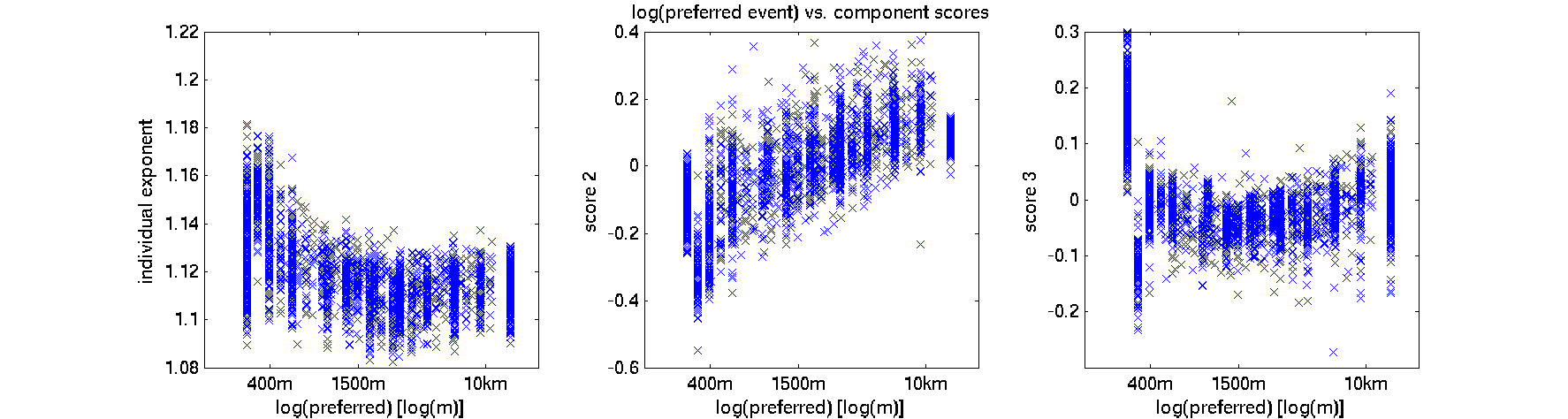}
\end{array}$
\end{center}
\caption{Scatter plots of training standard vs.~three-number-summary (top) and preferred distance vs.~three-number-summary.
In each case the individual exponents, 2nd and 3rd scores ($\lambda_2$, $\lambda_3$) are displayed on the $y$-axis and
the log-preferred distance and training standard on the $x$-axis.
\label{fig:meaning_components}}
\end{figure}

\noindent
{\bf (S.III.a) Further exploration of the three-number-summary.} Scatter plots of preferred distance and training standard against the athletes' three-number-summaries are displayed in Figure~\ref{fig:meaning_components}.
The training standard correlates predominantly with the individual exponent (score 1); score 1 vs. standard---$r=-0.89$ ($p \le 0.001$); score 2 vs. standard---$r=0.22$ ($p \le 0.001$); score 3 vs. standard---$r=0.031$ ($p = 0.07$);
all correlations are Spearman correlations with significance computed using a $t$-distribution approximation to the correlation coefficient under the null.
On the other hand preferred distance is associated with all three numbers in the summary, especially the second; score 1 vs. log(specialization)---$r=0.29$ ($p \le 0.001$); score 2 vs. log(specialization)---$r=-0.58$ ($p \le 0.001$); score 3 vs. log(specialization)---$r=-0.14$ ($p = \le 0.001$);
The association between the third score and specialization is non-linear with an optimal value around the middle distances. We stress that low correlation does not imply low predictive power; the whole summary should be considered as a whole, and the LMC predictor is non-linear. Also, we observe that correlations increase when considering only performances over certain distances, see Figure~\ref{fig:svs}.\\

\begin{figure}
\begin{center}
$\begin{array}{c}
\includegraphics[width=100mm,clip=true,trim= 0mm 0mm 0mm 0mm]{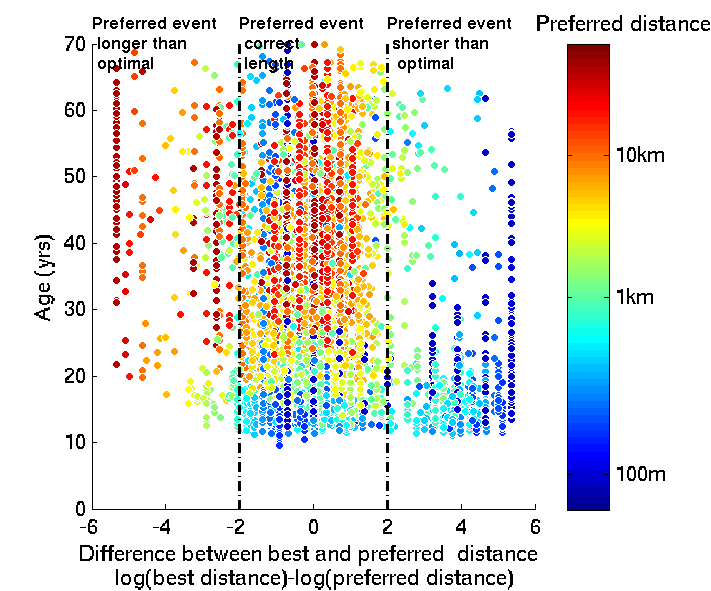}
\end{array}$
\end{center}
\caption{Difference of preferred distance and optimal distance, versus age of the athlete, colored by specialization distance. Most athletes prefer the distance they are predicted to be best at. There is a mismatch of best and preferred for a group of younger athletes who have greater potential over longer distances, and for a group of older athletes who's potential is maximized over shorter distances than attempted.
\label{fig:best_event_pred}}
\end{figure}

\noindent
{\bf (S.III.b) Preferred event vs best event.} For the top 95\% male athletes who have attempted 3 or more events, we use LMC rank 2 to compute which percentile they would achieve in each event. We then determine the distance of the event at which they would achieve the best percentile, to which we will refer as the ``optimal distance''. Figure~\ref{fig:best_event_pred} shows for each athlete the difference between their preferred and optimal distance.

It can be observed that the large majority of athletes prefer to attempt events in the vicinity of their optimal event. There is a group of young athletes who attempt events which are shorter than the predicted optimal distance, and a group of old athletes attempting events which are longer than optimal. One may hypothesize that both groups could be explained by social phenomena: young athletes usually start to train on shorter distances, regardless of their potential over long distances. Older athletes may be biased to attempting endurance type events. \\

\noindent
{\bf (S.IV) Pivoting and phase transitions.} We look more closely at the pivoting phenomenon illustrated in Figure~\ref{fig:illustration} top right, and the phase transition discussed in observation (V). We consider the top $25\%$ of male athletes who have attempted at least 3 events, in their best year.

We compute 10 performances of equivalent standard by using LMC in rank 1 in log-time coordinates, by
setting a benchmark performance over the marathon and sequentially predicting each lower distance (marathon
predicts HM, HM predicts 10km etc.). This yields equivalent benchmark performances $t_{1},\dots,t_{10}$.

We then consider triples of consecutive distances $s_{i-1},s_i,s_{i+1}$ (excluding the Mile since close in distance to the 1500m) and study the pivoting behaviour on the data set, by performing the analogous prediction displayed Figure~\ref{fig:illustration}.

More specifically, for each triple, we predict the performance on the distance $s_{i+1}$ using LMC rank 2, from the performances over the distances $s_{i-1}$ and $s_{i}$. The prediction is performed in two ways, once with and once without perturbation of the benchmark performance at $s_{i-1}$, which we then compare. Intuitively, this corresponds to comparing the red to the green curve in Figure~\ref{fig:illustration}. In mathematical terms:

\begin{enumerate}
\item We obtain a prediction $\widehat{t}_{i+1}$ for the distance $s_{i+1}$ from the benchmark performances $t_{i}$, $t_{i-1}$ and consider this as the unperturbed prediction, and
\item We obtain a prediction $\widehat{t}_{i+1}+\delta(\epsilon)$ for the distance $s_{i+1}$ from the benchmark performance $t_{i}$ on $s_i$ and the perturbed performance $(1+\epsilon)t_{i-1}$ on the distance $s_{i-1}$, considering this as the perturbed prediction.
\end{enumerate}

\begin{figure}
\begin{center}
$\begin{array}{c}
\includegraphics[width=100mm,clip=true,trim= 0mm 0mm 0mm 0mm]{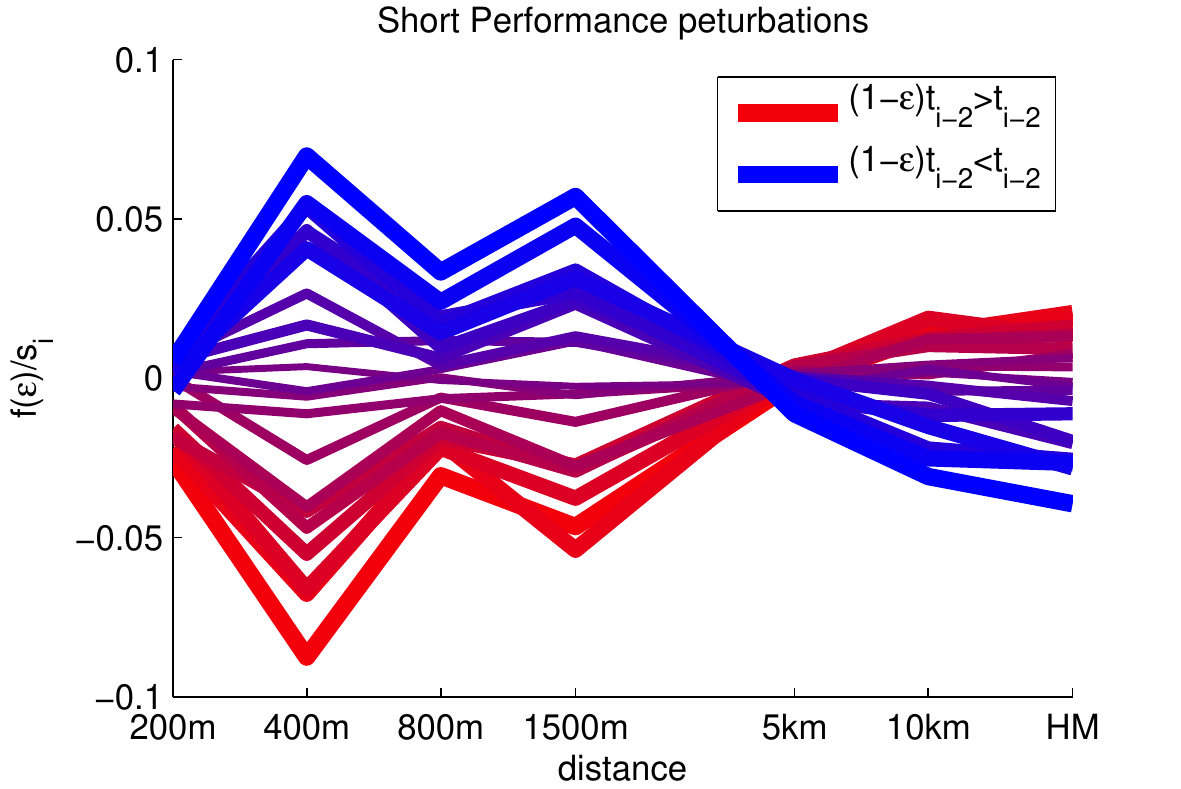}
\end{array}$
\end{center}
\caption{Pivot phenomenon in the low-rank model. The figure quantifies the strength and sign of pivoting as in Figure~\ref{fig:illustration}, top right, at different middle distances $s_i$ (x-axis). The computations are based on equivalent log-time performances $t_{i-1},t_i,t_{i+1}$ at consecutive triples $s_{i-1},s_i,s_{i+1}$ of distances. The y-coordinate indicates the signed relative change of the LMC rank 2 prediction of $t_{i+1}$ from $t_{i-1}$ and $t_i$ changes, when $t_i$ is fixed and $t_{i-1}$ undergoes a relative change of $1\%, 2\%,\ldots, 10\%$ (red curves, line thickness is proportional to change), or $-1\%, -2\%,\ldots, -10\%$ (blue curves, line thickness is proportional to change).
For example, the largest peak corresponds to a middle distance of $s_i = 400m$. When predicting 800m from 400m and 200m, the predicted log-time $t_{i+1}$ (= 800m performance) decreases by 8\% when $t_{i-1}$ (= 200m performance) is increased by 10\% while $t_i$ (= 400m performance) is kept constant.
\label{fig:pivots}}
\end{figure}

We record these estimates for $\epsilon = -0.1,0.09,\dots,0,0.01,\dots,0.1$ and calculate the relative change of the perturbed prediction with respect to the unperturbed, which is $\delta_i(\epsilon)/\widehat{t}_i$. The results are displayed in Figure~\ref{fig:pivots}.

We find that for pivot distances $s_{i}$ shorter than 5km, a slower performance on the shorter distance $s_{i-2}$ leads to
a faster performance over the longer distance $s_{i}$, insofar as this is predicted by the rank 2 predictor.
On the other hand we find that for pivot distances greater than or equal to 5km, a faster performance over the shorter distance
also implies a faster performance over the longer distance.

\noindent

\normalsize

\pagestyle{empty}
\bibliographystyle{plain}
{\small
\bibliography{runners}}

\begin{table}
\tiny
\begin{center}
\begin{tabular}{cccc|cc|cccc|cc|cc}
&&&\multicolumn{1}{c}{}& \multicolumn{2}{c}{\vtop{\hbox{\strut Generic}\hbox{\strut Baselines}}} & \multicolumn{4}{c}{\vtop{\hbox{\strut State of art}\hbox{\strut Performance Predictors}}} & \multicolumn{2}{c}{\vtop{\hbox{\strut State of art}\hbox{\strut Matrix Completion}}} & \multicolumn{2}{c}{\vtop{\hbox{\strut Proposed}\hbox{\strut Method: LMC}}} \\ 
\rotatebox{90}{evaluation} & \rotatebox{90}{percentiles} & \rotatebox{90}{no.events} & \rotatebox{90}{data type} & \rotatebox{90}{r.mean} & \rotatebox{90}{k-NN} & \rotatebox{90}{\parbox{1cm}{individual\\power law}} & \rotatebox{90}{riegel} & \rotatebox{90}{power law} & \rotatebox{90}{purdy} & \rotatebox{90}{\parbox{1cm}{nuclear\\norm}} & \rotatebox{90}{EM} & \rotatebox{90}{\parbox{1cm}{LMC\\rank 1}} & \rotatebox{90}{\parbox{1cm}{LMC\\rank 2}} \\ \hline 
log time & 0-95 & 3 & best & 0.131 & 0.122 & 0.103 & 0.0982 & 0.0973 & 0.061 & 0.391 & 0.0566 & 0.0586 & \bf 0.0515 \\ 
&&& & $\pm$0.003 & $\pm$0.003 & $\pm$0.004 & $\pm$0.005 & $\pm$0.005 & $\pm$0.003 & $\pm$0.05 & $\pm$0.003 & $\pm$0.003 & \bf $\pm$0.003 \\ 
normalized & 0-95 & 3 & best & 0.136 & 0.129 & 0.107 & 0.106 & 0.105 & 0.0684 & 0.19 & 0.0634 & 0.0643 & \bf 0.0576 \\ 
&&& & $\pm$0.004 & $\pm$0.005 & $\pm$0.005 & $\pm$0.007 & $\pm$0.007 & $\pm$0.004 & $\pm$0.01 & $\pm$0.005 & $\pm$0.004 & \bf $\pm$0.004 \\ 
speed & 0-95 & 3 & best & 0.666 & 0.614 & 0.61 & 0.547 & 0.541 & 0.308 & 26.6 & 0.292 & 0.312 & \bf 0.253 \\ 
&&& & $\pm$0.01 & $\pm$0.01 & $\pm$0.02 & $\pm$0.03 & $\pm$0.02 & $\pm$0.02 & $\pm$10 & $\pm$0.02 & $\pm$0.01 & \bf $\pm$0.01 \\ 
log time & 0-95 & 3 & random & 0.138 & 0.126 & 0.0931 & 0.0931 & 0.0919 & 0.0591 & 0.442 & 0.0561 & 0.0567 & \bf 0.0471 \\ 
&&& & $\pm$0.003 & $\pm$0.003 & $\pm$0.004 & $\pm$0.004 & $\pm$0.004 & $\pm$0.003 & $\pm$0.04 & $\pm$0.003 & $\pm$0.003 & \bf $\pm$0.002 \\ 
normalized & 0-95 & 3 & random & 0.145 & 0.134 & 0.0951 & 0.101 & 0.0998 & 0.0682 & 0.205 & 0.0634 & 0.064 & \bf 0.0538 \\ 
&&& & $\pm$0.004 & $\pm$0.004 & $\pm$0.004 & $\pm$0.005 & $\pm$0.005 & $\pm$0.004 & $\pm$0.01 & $\pm$0.004 & $\pm$0.004 & \bf $\pm$0.004 \\ 
speed & 0-95 & 3 & random & 0.693 & 0.617 & 0.592 & 0.505 & 0.498 & 0.284 & 24.7 & 0.28 & 0.286 & \bf 0.226 \\ 
&&& & $\pm$0.01 & $\pm$0.01 & $\pm$0.03 & $\pm$0.02 & $\pm$0.02 & $\pm$0.01 & $\pm$10 & $\pm$0.02 & $\pm$0.01 & \bf $\pm$0.01 \\ 
log time & 0-95 & 4 & best & 0.127 & 0.12 & 0.0777 & 0.0819 & 0.0822 & 0.0581 & 0.178 & 0.0529 & 0.0536 & \bf 0.0467 \\ 
&&& & $\pm$0.003 & $\pm$0.003 & $\pm$0.002 & $\pm$0.003 & $\pm$0.003 & $\pm$0.002 & $\pm$0.02 & $\pm$0.002 & $\pm$0.002 & \bf $\pm$0.002 \\ 
log time & 0-25 & 3 & best & 0.0557 & 0.0528 & 0.0806 & 0.0683 & 0.072 & 0.0411 & 0.301 & 0.0383 & 0.0411 & \bf 0.0306 \\ 
&&& & $\pm$0.001 & $\pm$0.001 & $\pm$0.003 & $\pm$0.003 & $\pm$0.003 & $\pm$0.001 & $\pm$0.03 & $\pm$0.001 & $\pm$0.001 & \bf $\pm$0.001 \\ 
\end{tabular}
\end{center}
\caption{Out-of-sample RMSE for prediction methods on different data setups. Predicted performance is of the 25 top percentiles of male athletes, in their best year. Standard errors are bootstrap estimates over 1000 repetitions. Compared method classes are (1) generic baselines, (2) state-of-the-art in performance prediction, (3) state-of-the-art in matrix completion, (4) local matrix completion (columns). Methods are (1.a) r.mean: predicting the mean of all athletes (1.b) k-NN: predicting the nearest neighbor. (2.a) riegel: Riegel's formula (2.b) power-law: power law with free exponent and coefficient. Exponent is the same for all athletes. (2.c) ind.power-law: power law with free exponent and coefficient. (2.d) purdy: Purdy points scheme (3.a) EM: expectation maximization (3.b) nuclear norm: nuclear norm minimization (4.a) LMC with rank 1 (4.b) LMC with rank 2. Data setup is specified by (i) evaluation: what is predicted. log-time = natural logarithm of time in seconds, normalized = time relative to mean performance, speed = average speed in meters per seconds, (ii) percentiles: selected percentile range of athletes, (iii) no.events tried = sub-set of athletes who have attempted at least that number of different events, (iv) data type: collation mode of performance matrix; best = 1 year around best performance, random = random 2 year period.
LMC rank 2 significantly outperforms all competitors in either setting.
\label{tab:compare_methods}}
\end{table}

\begin{table}
\tiny
\begin{center}
\begin{tabular}{cccc|cc|cccc|cc|cc}
&&&\multicolumn{1}{c}{}& \multicolumn{2}{c}{\vtop{\hbox{\strut Generic}\hbox{\strut Baselines}}} & \multicolumn{4}{c}{\vtop{\hbox{\strut State of art}\hbox{\strut Performance Predictors}}} & \multicolumn{2}{c}{\vtop{\hbox{\strut State of art}\hbox{\strut Matrix Completion}}} & \multicolumn{2}{c}{\vtop{\hbox{\strut Proposed}\hbox{\strut Method: LMC}}} \\ 
\rotatebox{90}{evaluation} & \rotatebox{90}{percentiles} & \rotatebox{90}{no.events} & \rotatebox{90}{data type} & \rotatebox{90}{r.mean} & \rotatebox{90}{k-NN} & \rotatebox{90}{\parbox{1cm}{individual\\power law}} & \rotatebox{90}{riegel} & \rotatebox{90}{power law} & \rotatebox{90}{purdy} & \rotatebox{90}{\parbox{1cm}{nuclear\\norm}} & \rotatebox{90}{EM} & \rotatebox{90}{\parbox{1cm}{LMC\\rank 1}} & \rotatebox{90}{\parbox{1cm}{LMC\\rank 2}} \\ \hline 
log time & 0-95 & 3 & best & 0.105 & 0.0962 & 0.0696 & 0.0661 & 0.0654 & 0.0423 & 0.128 & 0.0387 & 0.0402 & \bf 0.0336 \\ 
&&& & $\pm$0.002 & $\pm$0.002 & $\pm$0.002 & $\pm$0.002 & $\pm$0.002 & $\pm$0.001 & $\pm$0.01 & $\pm$0.001 & $\pm$0.001 & \bf $\pm$0.001 \\ 
normalized & 0-95 & 3 & best & 0.106 & 0.0972 & 0.07 & 0.0681 & 0.0674 & 0.0441 & 0.0907 & 0.04 & 0.0413 & \bf 0.0347 \\ 
&&& & $\pm$0.003 & $\pm$0.002 & $\pm$0.003 & $\pm$0.003 & $\pm$0.003 & $\pm$0.002 & $\pm$0.005 & $\pm$0.002 & $\pm$0.002 & \bf $\pm$0.001 \\ 
speed & 0-95 & 3 & best & 0.546 & 0.497 & 0.399 & 0.364 & 0.36 & 0.22 & 2.28 & 0.202 & 0.213 & \bf 0.172 \\ 
&&& & $\pm$0.01 & $\pm$0.01 & $\pm$0.01 & $\pm$0.01 & $\pm$0.01 & $\pm$0.007 & $\pm$0.8 & $\pm$0.007 & $\pm$0.007 & \bf $\pm$0.006 \\ 
log time & 0-95 & 3 & random & 0.112 & 0.0996 & 0.0655 & 0.0656 & 0.0646 & 0.0411 & 0.15 & 0.0376 & 0.0385 & \bf 0.032 \\ 
&&& & $\pm$0.003 & $\pm$0.002 & $\pm$0.002 & $\pm$0.002 & $\pm$0.002 & $\pm$0.001 & $\pm$0.01 & $\pm$0.001 & $\pm$0.001 & \bf $\pm$0.001 \\ 
normalized & 0-95 & 3 & random & 0.114 & 0.101 & 0.066 & 0.0686 & 0.0676 & 0.0438 & 0.101 & 0.0395 & 0.0405 & \bf 0.0338 \\ 
&&& & $\pm$0.003 & $\pm$0.003 & $\pm$0.002 & $\pm$0.002 & $\pm$0.002 & $\pm$0.002 & $\pm$0.006 & $\pm$0.002 & $\pm$0.002 & \bf $\pm$0.001 \\ 
speed & 0-95 & 3 & random & 0.573 & 0.505 & 0.38 & 0.355 & 0.35 & 0.207 & 2.56 & 0.191 & 0.2 & \bf 0.161 \\ 
&&& & $\pm$0.01 & $\pm$0.01 & $\pm$0.01 & $\pm$0.01 & $\pm$0.01 & $\pm$0.006 & $\pm$0.8 & $\pm$0.007 & $\pm$0.007 & \bf $\pm$0.005 \\ 
log time & 0-95 & 4 & best & 0.101 & 0.0958 & 0.0557 & 0.0551 & 0.0553 & 0.0406 & 0.0716 & 0.035 & 0.0366 & \bf 0.031 \\ 
&&& & $\pm$0.002 & $\pm$0.002 & $\pm$0.002 & $\pm$0.002 & $\pm$0.002 & $\pm$0.001 & $\pm$0.005 & $\pm$0.001 & $\pm$0.001 & \bf $\pm$0.001 \\ 
log time & 0-25 & 3 & best & 0.0424 & 0.0409 & 0.0559 & 0.0479 & 0.0507 & 0.031 & 0.097 & 0.0282 & 0.03 & \bf 0.0221 \\ 
&&& & $\pm$0.001 & $\pm$0.001 & $\pm$0.002 & $\pm$0.002 & $\pm$0.002 & $\pm$0.0008 & $\pm$0.009 & $\pm$0.0008 & $\pm$0.0009 & \bf $\pm$0.0007 \\ 
\end{tabular}
\end{center}
\caption{Out-of-sample MAE for prediction methods on different data setups. Predicted performance is of the 25 top percentiles of male athletes, in their best year. Standard errors are bootstrap estimates over 1000 repetitions. Compared method classes are (1) generic baselines, (2) state-of-the-art in performance prediction, (3) state-of-the-art in matrix completion, (4) local matrix completion (columns). Methods are (1.a) r.mean: predicting the mean of all athletes (1.b) k-NN: predicting the nearest neighbor. (2.a) riegel: Riegel's formula (2.b) power-law: power law with free exponent and coefficient. Exponent is the same for all athletes. (2.c) ind.power-law: power law with free exponent and coefficient. (2.d) purdy: Purdy points scheme (3.a) EM: expectation maximization (3.b) nuclear norm: nuclear norm minimization (4.a) LMC with rank 1 (4.b) LMC with rank 2. Data setup is specified by (i) evaluation: what is predicted. log-time = natural logarithm of time in seconds, normalized = time relative to mean performance, speed = average speed in meters per seconds, (ii) percentiles: selected percentile range of athletes, (iii) no.events tried = sub-set of athletes who have attempted at least that number of different events, (iv) data type: collation mode of performance matrix; best = 1 year around best performance, random = random 2 year period.
LMC rank 2 significantly outperforms all competitors in either setting.
\label{tab:compare_methods_mae}}
\end{table}

\begin{table}
\tiny
\begin{center}
\begin{tabular}{cccc|cc|cccc|cc|cc}
&&&\multicolumn{1}{c}{}& \multicolumn{2}{c}{\vtop{\hbox{\strut Generic}\hbox{\strut Baselines}}} & \multicolumn{4}{c}{\vtop{\hbox{\strut State of art}\hbox{\strut Performance Predictors}}} & \multicolumn{2}{c}{\vtop{\hbox{\strut State of art}\hbox{\strut Matrix Completion}}} & \multicolumn{2}{c}{\vtop{\hbox{\strut Proposed}\hbox{\strut Method: LMC}}} \\ 
\rotatebox{90}{evaluation} & \rotatebox{90}{percentiles} & \rotatebox{90}{no.events} & \rotatebox{90}{data type} & \rotatebox{90}{r.mean} & \rotatebox{90}{k-NN} & \rotatebox{90}{\parbox{1cm}{individual\\power law}} & \rotatebox{90}{riegel} & \rotatebox{90}{power law} & \rotatebox{90}{purdy} & \rotatebox{90}{\parbox{1cm}{nuclear\\norm}} & \rotatebox{90}{EM} & \rotatebox{90}{\parbox{1cm}{LMC\\rank 1}} & \rotatebox{90}{\parbox{1cm}{LMC\\rank 2}} \\ \hline 
log time & 0-95 & 3 & best & 0.14 & 0.13 & 0.106 & 0.1 & 0.0991 & 0.0639 & 0.362 & 0.0574 & 0.0622 & \bf 0.0545 \\ 
&&& & $\pm$0.007 & $\pm$0.006 & $\pm$0.005 & $\pm$0.006 & $\pm$0.006 & $\pm$0.007 & $\pm$0.08 & $\pm$0.006 & $\pm$0.005 & \bf $\pm$0.005 \\ 
normalized & 0-95 & 3 & best & 0.148 & 0.139 & 0.11 & 0.105 & 0.104 & 0.0724 & 0.177 & 0.0658 & 0.0694 & \bf 0.062 \\ 
&&& & $\pm$0.009 & $\pm$0.009 & $\pm$0.007 & $\pm$0.007 & $\pm$0.007 & $\pm$0.01 & $\pm$0.02 & $\pm$0.009 & $\pm$0.008 & \bf $\pm$0.008 \\ 
speed & 0-95 & 3 & best & 0.715 & 0.655 & 0.655 & 0.583 & 0.577 & 0.338 & 19.2 & 0.307 & 0.341 & \bf 0.292 \\ 
&&& & $\pm$0.03 & $\pm$0.03 & $\pm$0.04 & $\pm$0.04 & $\pm$0.04 & $\pm$0.04 & $\pm$10 & $\pm$0.04 & $\pm$0.03 & \bf $\pm$0.03 \\ 
log time & 0-95 & 3 & random & 0.138 & 0.126 & 0.0931 & 0.0931 & 0.0919 & 0.0591 & 0.442 & 0.0561 & 0.0567 & \bf 0.0471 \\ 
&&& & $\pm$0.003 & $\pm$0.003 & $\pm$0.004 & $\pm$0.004 & $\pm$0.004 & $\pm$0.003 & $\pm$0.04 & $\pm$0.003 & $\pm$0.003 & \bf $\pm$0.002 \\ 
normalized & 0-95 & 3 & random & 0.145 & 0.134 & 0.0951 & 0.101 & 0.0998 & 0.0682 & 0.205 & 0.0634 & 0.064 & \bf 0.0538 \\ 
&&& & $\pm$0.004 & $\pm$0.004 & $\pm$0.004 & $\pm$0.005 & $\pm$0.005 & $\pm$0.004 & $\pm$0.01 & $\pm$0.004 & $\pm$0.004 & \bf $\pm$0.003 \\ 
speed & 0-95 & 3 & random & 0.693 & 0.617 & 0.592 & 0.505 & 0.498 & 0.284 & 24.7 & 0.28 & 0.286 & \bf 0.226 \\ 
&&& & $\pm$0.01 & $\pm$0.01 & $\pm$0.03 & $\pm$0.02 & $\pm$0.02 & $\pm$0.01 & $\pm$10 & $\pm$0.02 & $\pm$0.01 & \bf $\pm$0.01 \\ 
log time & 0-95 & 4 & best & 0.137 & 0.128 & 0.0823 & 0.0859 & 0.0862 & 0.062 & 0.237 & 0.0608 & 0.0599 & \bf 0.0531 \\ 
&&& & $\pm$0.007 & $\pm$0.007 & $\pm$0.004 & $\pm$0.006 & $\pm$0.006 & $\pm$0.004 & $\pm$0.04 & $\pm$0.006 & $\pm$0.004 & \bf $\pm$0.004 \\ 
log time & 0-25 & 3 & best & 0.0539 & 0.052 & 0.081 & 0.0675 & 0.071 & 0.0412 & 0.248 & 0.0358 & 0.0417 & \bf 0.0318 \\ 
&&& & $\pm$0.003 & $\pm$0.003 & $\pm$0.005 & $\pm$0.005 & $\pm$0.005 & $\pm$0.003 & $\pm$0.06 & $\pm$0.002 & $\pm$0.003 & \bf $\pm$0.002 \\ 
\end{tabular}
\end{center}
\caption{Prediction only from events which are earlier in time than the performance to be predicted. The table shows out-of-sample RMSE for performance prediction methods on different data setups. Predicted performance is of the 25 top percentiles of male athletes, in their best year. Standard errors are bootstrap estimates over 1000 repetitions. Legend is as in Table~\ref{tab:compare_methods}.
\label{tab:rrmse_causal}}
\end{table}

\begin{table}
\tiny
\begin{center}
\begin{tabular}{cccc|cc|cccc|cc|cc}
&&&\multicolumn{1}{c}{}& \multicolumn{2}{c}{\vtop{\hbox{\strut Generic}\hbox{\strut Baselines}}} & \multicolumn{4}{c}{\vtop{\hbox{\strut State of art}\hbox{\strut Performance Predictors}}} & \multicolumn{2}{c}{\vtop{\hbox{\strut State of art}\hbox{\strut Matrix Completion}}} & \multicolumn{2}{c}{\vtop{\hbox{\strut Proposed}\hbox{\strut Method: LMC}}} \\ 
\rotatebox{90}{evaluation} & \rotatebox{90}{percentiles} & \rotatebox{90}{no.events} & \rotatebox{90}{data type} & \rotatebox{90}{r.mean} & \rotatebox{90}{k-NN} & \rotatebox{90}{\parbox{1cm}{individual\\power law}} & \rotatebox{90}{riegel} & \rotatebox{90}{power law} & \rotatebox{90}{purdy} & \rotatebox{90}{\parbox{1cm}{nuclear\\norm}} & \rotatebox{90}{EM} & \rotatebox{90}{\parbox{1cm}{LMC\\rank 1}} & \rotatebox{90}{\parbox{1cm}{LMC\\rank 2}} \\ \hline 
time & 0-95 & 3 & best & 0.13 & 0.119 & 0.0959 & 0.0973 & 0.0964 & 0.0596 & 0.178 & 0.056 & 0.0569 & \bf 0.0499 \\ 
&&& & $\pm$0.003 & $\pm$0.003 & $\pm$0.003 & $\pm$0.006 & $\pm$0.006 & $\pm$0.003 & $\pm$0.01 & $\pm$0.003 & $\pm$0.002 & \bf $\pm$0.002 \\ 
time & 0-95 & 3 & random & 0.136 & 0.121 & 0.0874 & 0.0907 & 0.0895 & 0.0585 & 0.196 & 0.0544 & 0.055 & \bf 0.0461 \\ 
&&& & $\pm$0.003 & $\pm$0.003 & $\pm$0.003 & $\pm$0.003 & $\pm$0.003 & $\pm$0.003 & $\pm$0.01 & $\pm$0.002 & $\pm$0.002 & \bf $\pm$0.002 \\ 
time & 0-95 & 4 & best & 0.123 & 0.118 & 0.075 & 0.0782 & 0.0785 & 0.0566 & 0.117 & 0.0525 & 0.0522 & \bf 0.0455 \\ 
&&& & $\pm$0.003 & $\pm$0.003 & $\pm$0.002 & $\pm$0.003 & $\pm$0.003 & $\pm$0.002 & $\pm$0.008 & $\pm$0.003 & $\pm$0.002 & \bf $\pm$0.002 \\ 
time & 0-25 & 3 & best & 0.0559 & 0.053 & 0.076 & 0.0668 & 0.0704 & 0.0406 & 0.158 & 0.0377 & 0.0402 & \bf 0.0302 \\ 
&&& & $\pm$0.001 & $\pm$0.001 & $\pm$0.003 & $\pm$0.002 & $\pm$0.002 & $\pm$0.001 & $\pm$0.01 & $\pm$0.001 & $\pm$0.001 & \bf $\pm$0.001 \\ 
\end{tabular}
\end{center}
\caption{ Exactly the same table as Table~\ref{tab:compare_methods} but relative root mean squared errors reported in terms of
time. Models are learnt on the performances in log-time.
\label{tab:rrmse_time}}
\end{table}

\begin{table}
\tiny
\begin{center}
\begin{tabular}{cccc|cc|cccc|cc|cc}
&&&\multicolumn{1}{c}{}& \multicolumn{2}{c}{\vtop{\hbox{\strut Generic}\hbox{\strut Baselines}}} & \multicolumn{4}{c}{\vtop{\hbox{\strut State of art}\hbox{\strut Performance Predictors}}} & \multicolumn{2}{c}{\vtop{\hbox{\strut State of art}\hbox{\strut Matrix Completion}}} & \multicolumn{2}{c}{\vtop{\hbox{\strut Proposed}\hbox{\strut Method: LMC}}} \\ 
\rotatebox{90}{evaluation} & \rotatebox{90}{percentiles} & \rotatebox{90}{no.events} & \rotatebox{90}{data type} & \rotatebox{90}{r.mean} & \rotatebox{90}{k-NN} & \rotatebox{90}{\parbox{1cm}{individual\\power law}} & \rotatebox{90}{riegel} & \rotatebox{90}{power law} & \rotatebox{90}{purdy} & \rotatebox{90}{\parbox{1cm}{nuclear\\norm}} & \rotatebox{90}{EM} & \rotatebox{90}{\parbox{1cm}{LMC\\rank 1}} & \rotatebox{90}{\parbox{1cm}{LMC\\rank 2}} \\ \hline 
time & 0-95 & 3 & best & 0.106 & 0.0954 & 0.0669 & 0.0654 & 0.0647 & 0.042 & 0.0876 & 0.0384 & 0.0397 & \bf 0.0333 \\ 
&&& & $\pm$0.002 & $\pm$0.002 & $\pm$0.002 & $\pm$0.002 & $\pm$0.002 & $\pm$0.001 & $\pm$0.005 & $\pm$0.001 & $\pm$0.001 & \bf $\pm$0.001 \\ 
time & 0-95 & 3 & random & 0.112 & 0.0982 & 0.0635 & 0.0651 & 0.0642 & 0.041 & 0.098 & 0.0373 & 0.0381 & \bf 0.0318 \\ 
&&& & $\pm$0.002 & $\pm$0.002 & $\pm$0.002 & $\pm$0.002 & $\pm$0.002 & $\pm$0.001 & $\pm$0.006 & $\pm$0.001 & $\pm$0.001 & \bf $\pm$0.001 \\ 
time & 0-95 & 4 & best & 0.101 & 0.0954 & 0.0547 & 0.054 & 0.0543 & 0.0401 & 0.0605 & 0.0348 & 0.0362 & \bf 0.0307 \\ 
&&& & $\pm$0.002 & $\pm$0.002 & $\pm$0.002 & $\pm$0.002 & $\pm$0.002 & $\pm$0.001 & $\pm$0.003 & $\pm$0.001 & $\pm$0.001 & \bf $\pm$0.001 \\ 
time & 0-25 & 3 & best & 0.0425 & 0.041 & 0.0542 & 0.0476 & 0.0504 & 0.0308 & 0.0688 & 0.028 & 0.0297 & \bf 0.022 \\ 
&&& & $\pm$0.001 & $\pm$0.001 & $\pm$0.002 & $\pm$0.001 & $\pm$0.002 & $\pm$0.0008 & $\pm$0.005 & $\pm$0.0008 & $\pm$0.0009 & \bf $\pm$0.0006 \\ 
\end{tabular}
\end{center}
\caption{Exactly the same table as Table~\ref{tab:compare_methods} but relative mean absolute errors reported in terms of
time. Models are learnt on the performances in log-time.
\label{tab:rmae_time}}
\end{table}

\begin{table}
\scriptsize
\begin{center}
\begin{tabular}{c|cccc}
\rotatebox{90}{no events.} & \rotatebox{90}{r1} & \rotatebox{90}{r2} & \rotatebox{90}{r3} & \rotatebox{90}{r4} \\ \hline 
3 & 0.0411 & \bf 0.0306 & --- & --- \\ 
 & $\pm$0.001 & \bf $\pm$0.001 &  &  \\ 
4 & 0.0446 & 0.0328 & \bf 0.0309 & --- \\ 
 & $\pm$0.002 & $\pm$0.001 & \bf $\pm$0.001 &  \\ 
5 & 0.0518 & 0.0408 & \bf 0.04 & 0.0408 \\ 
 & $\pm$0.003 & $\pm$0.003 & \bf $\pm$0.003 & $\pm$0.004 \\ 
\end{tabular}
\end{center}
\caption{Determination of the true rank of the model. Table displays out-of-sample RMSE for predicting performance with LMC rank 1-4 (columns) Predicted performance is of the 25 top percentiles of male athletes, in their best year, who have attempted at least the number of events indicated by the row. The model is learnt on performances in log-time coordinates. Standard errors are bootstrap estimates over 1000 repetitions. The entries where {\bf no.~events} $\ge$ rank are empty, as LMC rank $r$ needs $r+1$ attempted events for leave-one-out-validation. Prediction with LMC rank $3$ is always better or equally good compared to using a different rank, in terms of out-of-sample prediction accuracy.
\label{tab:determine_rank}}
\end{table}

\begin{table}
\scriptsize
\begin{center}
\begin{tabular}{c|c}
\rotatebox{90}{subgroup} & \rotatebox{90}{RMSE} \\ \hline 
Amateur & \bf 0.0305 \\ 
 & \bf $\pm$0.0002 \\ 
Female & \bf 0.0305 \\ 
 & \bf $\pm$0.0003 \\ 
Old & \bf 0.0326 \\ 
 & \bf $\pm$0.0003 \\ 
\end{tabular}
\end{center}
\caption{Prediction in three different subgroups: amateur athletes, female athletes, older athletes. Table displays out-of-sample RMSE for predicting performance with LMC rank 2.
\label{tab:generalize}}
\end{table}

\begin{table}
\scriptsize
\begin{center}
\begin{tabular}{c|ccc}
\rotatebox{90}{rank} & \rotatebox{90}{log time} & \rotatebox{90}{speed} & \rotatebox{90}{normalized} \\ \hline 
 1  & 0.041 & 0.0376 & 0.0399 \\ 
 & $\pm$0.001 & $\pm$0.001 & $\pm$0.001 \\ 
 2  & 0.0304 & 0.0315 & 0.0305 \\ 
 & $\pm$0.001 & $\pm$0.001 & $\pm$0.001 \\ 
\end{tabular}
\end{center}
\caption{Effect of performance measure in which the LMC model is learnt. The model is learnt on three different measures of performance: log-time, time normalized by event mean, speed (columns). The table shows out-of-sample RMSE for predicting log-time performance with LMC rank 1,2. Standard errors are bootstrap estimates over 1000 repetitions. Performance is of the 25 top percentiles of male athletes, in their best year of performance.
\label{tab:choose_feature}}
\end{table}

\begin{table}
\scriptsize
\begin{center}
\begin{tabular}{cc|cccc}
\rotatebox{90}{percentiles} & \rotatebox{90}{no.event} & \rotatebox{90}{bagged LMC r2} & \rotatebox{90}{bagged power-law} & \rotatebox{90}{LMC r2} & \rotatebox{90}{power-law} \\ \hline 
0-25 & 3 & 0.031 & 0.0654 & \bf 0.0308 & 0.0666 \\ 
& & $\pm$0.001 & $\pm$0.002 & \bf $\pm$0.001 & $\pm$0.003 \\ 
0-95 & 3 & 0.0529 & 0.0898 & \bf 0.0512 & 0.0948 \\ 
& & $\pm$0.003 & $\pm$0.004 & \bf $\pm$0.003 & $\pm$0.004 \\ 
0-95 & 4 & 0.048 & 0.0762 & \bf 0.0467 & 0.0825 \\ 
& & $\pm$0.003 & $\pm$0.003 & \bf $\pm$0.002 & $\pm$0.003 \\ 
\end{tabular}
\end{center}
\caption{Comparison of prediction using all distances, to prediction using only closest distances. Table displayes out-of-sample RMSE of predicting log-time, for (5.a) the bagged power law and (5.b) the bagged LMC rank 2 predictor, compared with the un-bagged variants, (2.b) and (4.b). Predicted performance is of the 25 top percentiles of male athletes, in their best year. Standard errors are bootstrap estimates over 1000 repetitions. The results of the bagging predictors are very similar to the unbagged one.
\label{tab:learn_weights}}
\end{table}

\end{document}